\documentclass[onecolumn,amsmath,amssymb,nofootinbib,11pt]{article}
\pdfoutput=1
\usepackage{jheppub}
\usepackage{ifpdf}
\usepackage{bbold}
\usepackage{csquotes}

\usepackage{graphicx,subcaption}
\usepackage{amsmath}
\usepackage{amssymb}
\usepackage{epsfig}
\usepackage{pdfpages}
\usepackage{graphicx,epstopdf}
\usepackage[makeroom]{cancel}
\usepackage{hyperref}
\hypersetup{pdftex,colorlinks=true,allcolors=blue}
\usepackage{array}
\usepackage{tikz}
\usepackage{mathtools}
\usepackage{braket}
\usepackage{yhmath}
\usepackage[thinlines]{easytable}
\usepackage{enumitem}

\newcommand{\no}{\nonumber}
\newcommand{\der}{\partial}
\newcommand{\ie}{\textit{i.e.}}
\newcommand{\mC}{\mathcal{C}}
\newcommand{\be}{\begin{equation}}
\newcommand{\ee}{\end{equation}}
\newcommand{\normord}[1]{:\mathrel{#1}:}

\graphicspath{{./Figures/}}

\begin{document}

	\title{Circuit Complexity and 2D Bosonisation}
	\author{Dongsheng Ge, Giuseppe Policastro}
	\affiliation{Laboratoire de Physique de l'\'Ecole Normale Sup\'erieure, ENS, Universit\'e PSL, CNRS, \\Sorbonne Universit\'e, Universit\'e Paris-Diderot, Sorbonne Paris Cit\'e, \\
24 rue Lhomond, 75005 Paris, France}
	\emailAdd{dongsheng.ge@ens.fr}
	\emailAdd{giuseppe.policastro@ens.fr}
	\abstract{We consider the circuit complexity of free bosons and free fermions in 1+1 dimensions. Motivated by the results of \cite{Jefferson:2017sdb} and \cite{Hackl:2018ptj,Khan:2018rzm}  who found different behavior in the complexity of free bosons and fermions, in any dimension, we consider the 1+1 dimensional case where, thanks to the  bosonisation equivalence of the Hilbert spaces, we can consider the same state from both the bosonic and the fermionic perspectives. This allows us to study the dependence of the complexity on the choice of the set of gates, which explains the discrepancy. 
	We study the effect in two classes of states: i) bosonic-coherent / fermionic-gaussian states; ii) states that are both bosonic- and fermionic-gaussian. We consider the complexity relative to the ground state. In the first class, the different complexities can be related to each other by introducing a mode-dependent cost function in one of the descriptions. The differences in the second class are more important, in terms of the structure of UV divergencies and the overall behavior of the complexity. 
	}
	\maketitle
	
\section{Introduction}

Recently, much attention has been devoted to the study of quantum complexity in connection with the holographic correspondence. 
Quantum complexity is a concept that has its origin in quantum computation theory as a means of characterizing the difficulty (in the sense of the amount of resources needed) of performing a task on a quantum computer; more precisely,  if the task can be described as producing a certain quantum state $\ket{\psi_T}$ from a given initial state $\ket{\psi_R}$ using a circuit made of elementary unitary operations (gates), the quantum complexity can be defined as the minimum number of gates required for such a circuit. 
It has been suggested by Susskind and collaborators \cite{Susskind:2014rva,Stanford:2014jda,Susskind:2014jwa,Susskind:2014moa} that the notion of complexity may be an important component in our understanding of the properties of emergent spacetime, and possibly 
point to a solution of the information loss paradox, by giving additional insight, beyond what can be obtained from entanglement entropy, into the information-theoretic properties associated to spacetime and in particular to the region inside black hole horizons (see the recent lectures \cite{Susskind:2018pmk}).  

It was further conjectured that, similarly to  entanglement entropy which is described holographically by  the area of a minimal surface \cite{Ryu:2006bv}, quantum complexity is also captured, in a theory with a holographic dual, by a 
simple gravitational observable. However what precisely is the observable is unclear. There are two different proposals: one takes the volume of a maximal spatial slice of the geometry  \cite{Stanford:2014jda}, the other takes the action evaluated on the \textit{Wheeler-DeWitt} patch \cite{Brown:2015bva,Brown:2015lvg}. Each proposal has its merits and drawbacks, and a clear-cut way to choose one over the other has 
not been found yet. In particular, both prescription lead to similar behavior for the late-time growth of complexity during the formation of a black-hole (i.e., linear growth in time, albeit with different coefficients \cite{Chapman:2018dem,Chapman:2018lsv}), and have similar structure of UV divergence 
\begin{equation} 
\mathcal{C} \sim a \frac{V}{\delta^{d-1}} (1+ \cal{O}(\delta)) \,,
\end{equation}
where $d$ is the spacetime dimension of the dual field theory, $\delta$ is a short-distance cutoff, and $a$ 
is  a coefficient that depends on the prescription.\footnote{In the case of the action, there can be a further divergence of the form $\frac{1}{\delta^{d-1}} \ln \delta$ \cite{Carmi:2016wjl}, but it can be removed by a boundary counterterm \cite{Lehner:2016vdi} which is also needed to make the prescription reparametrization invariant.}

Even in cases when the two prescriptions give different results (e.g., in the case of an $AdS_3$ space with a defect brane, as we found in our paper with S. Chapman \cite{Chapman:2018bqj}), we do not have at present a criterion for choosing one over the other, absent any independent calculation that can serve as benchmark. By contrast, in the case of the entanglement entropy, one can compute it in a 
2d CFT, for an interval of length $\ell$, and obtain the famous exact result \cite{Holzhey:1994we} 
\begin{equation}
S_{EE}= {c \over 3} \log \left({\ell \over \delta} \right) \,,
\end{equation}  
which depend only on the central charge $c$. Because the coefficient of the log does not change under a rescaling of the cutoff, it can be considered a universal quantity with a well-defined physical meaning. By analogy, one could try to attribute a similar universal meaning to the coefficient of the log in the complexity, but the situation is less clear. 

%
The problem is that the definition in terms of gates, which is applicable to a finite quantum system, does not translate easily to a definition that is applicable to a continuum quantum field theory. In other words, we do not know how to associate the notion of complexity to a well-defined observable in QFT. 
This problem was considered in 
\cite{Jefferson:2017sdb,Chapman:2017rqy}, who provided a partial answer by proposing a definition of complexity for free scalar fields. The proposal of \cite{Jefferson:2017sdb} used the Nielsen's approach of {\it geometrization} of quantum computation \cite{2007quant.ph..1004D}, while \cite{Chapman:2017rqy} used a method based on the Fubini-Study metric \footnote{For other approaches and follow-ups see \cite{Caputa:2017yrh,Bhattacharyya:2018wym,Roberts:2016hpo,Chapman:2018hou,Bhattacharyya:2018bbv,Belin:2018bpg,Yang:2018nda,Yang:2018tpo,Ali:2018aon}.}. 
In Nielsen's method, described in more detail in Section \ref{sec:nielsenreview}, the circuit is replaced by a continuous version that is a path in the space of unitary operators, of the form 
\begin{equation} \label{unitary-path}
U(s) = \mathcal{P}\textrm{exp} \left( -i \int_0^s ds' \, Y^I(s') \mathcal{O}_I \right) \,,
\end{equation} 
The functions $Y^I(s)$ are determined by minimization of a certain functional $F[Y^I(s)]$ that determines the {\it cost} associated to a given path. The complexity of a unitary operator $V$ is then defined as the minimum of the cost functional over all paths \eqref{unitary-path} such that $U(s=1) = V$. 

What is important to notice here is that this definition of complexity depends on several choices: the choice of the allowed space of operators used to build the circuit \footnote{If the space of all operators is allowed, then obviously all circuits would have the same complexity.}, and indeed of a specific basis $\mathcal{O}_I$,  the choice of a cost functional, and a choice of parametrization of the path.

The space of unitary operators acting on the Hilbert space of a QFT is very large, but one of the main points of \cite{Jefferson:2017sdb} was to reduce it to a tractable setup by considering operators that 
act within the subset of Gaussian states. This allows one to consider the complexity of any state that is the ground state of any Hamiltonian quadratic in the fields. 
Discretizing the free scalar field to a set of $N$ harmonic oscillators, a Gaussian state is described by a symmetric $2N\times 2N$ matrix, and the 
group of unitary operators that preserves the Gaussian states is $Sp(2N)$. In \cite{Jefferson:2017sdb} the behaviour of the complexity between Gaussian states was investigated for various choices of cost functions, mostly focusing on the one corresponding to the Cartan-Killing metric on the symplectic group. 

The approach used in \cite{Chapman:2017rqy} considers a path in the space of states; the complexity is identified with the minimal length of a path connecting two states, calculated in the standard Fubini-Study metric on the space of normalized states. For a parametrized path of quantum states $| \psi(\sigma) \rangle$, the line element is 
\begin{equation} 
ds = d\sigma \sqrt { \braket{ \der_\sigma \psi(\sigma) | \der_\sigma \psi(\sigma) }
	 - |\braket{\psi(\sigma) | \der_\sigma \psi(\sigma) }|^2} \,.
\end{equation}
This definition may seem more canonical, but if one allows the most general path in the space of states, then the geodesic distance between normalized states is always less or equal to $\pi/2$.  
In order to have a sensible measure of complexity, one must somehow restrict the possible paths. 
The proposal of  \cite{Chapman:2017rqy} is to use paths that can be obtained using unitary operators similar to \eqref{unitary-path}, with the basis operators being a subset of the 
 bilinears  the creation and annihilation operators; thus in both approaches one does not leave the space of Gaussian states, and the results are comparable. 


Nielsen's approach was extended to the case of free fermions in \cite{Hackl:2018ptj,Khan:2018rzm}. In comparison to the bosonic case, the main difference is that the relevant group of operators acting on 
Gaussian states is $SO(2N)$, which is a compact group, with the consequence that the complexity (again measured using the Cartan-Killing metric) cannot grow very large. 
Considering a field theory in $d$ spacetime dimension with spatial volume $V$ and a UV cutoff $\Lambda$, it turns out that the leading divergent term in the bosonic complexity is

\begin{equation}\label{boson-div}
\mathcal{C}_{\Lambda}^{b} \sim V \Lambda^{d-1} |\ln (\Lambda/\omega_0)|^\kappa,
\end{equation}
%
where $\omega_0$ is an arbitrary reference scale and $\kappa$ is a parameter related to the choice of the cost function, 
whereas for a  free fermion field theory is
\begin{equation}\label{fermion-div}
\mathcal{C}_{\Lambda}^{f} \sim V \Lambda^{d-1}. 
\end{equation}

The discrepancy seems at odds with the holographic interpretation, as the holographic result is relatively blind to the fermionic or bosonic nature of the dual fields. Moreover, for the case of a two-dimensional theory, we might think that the bosonization map between bosons and fermions should imply the equality of the two results. 
These observations do not constitute a strong objection to the results of \cite{Jefferson:2017sdb,Hackl:2018ptj,Chapman:2017rqy,Khan:2018rzm}, because states with a holographic dual are not Gaussian states, the theory is strongly coupled whereas \eqref{boson-div} and \eqref{fermion-div} are obtained for free fields, and  complexity is not a well-defined 
field theory observable in the usual sense, as we stressed. Nevertheless, they provide the motivation for the present paper, where we look more closely at the complexity for bosons and fermions 
in 1+1 dimensions, comparing states that are related by bosonisation. From this point of view, the results \eqref{boson-div} and \eqref{fermion-div} do not constitute a discrepancy, rather they 
illustrate the effect of a different choice of operator spaces. 

In the context of quantum computation, the question of the choice of gates is of obvious importance, and there are some known results: when the gates act on arrays of qubits, it is possible to show that there is a universal finite set of gates, such that any unitary operator can be approximated by a circuit made of universal gates with an arbitrarily small error $\epsilon$ (see {\it e.g.} \cite{2008arXiv0804.3401W}), and the size of the circuit is  ${\cal O}(\ln^c(1/\epsilon))$, for some constant $c$. Moreover, if a circuit has complexity $m$ with some choice of basic gates, a different choice will give complexity ${\cal O}(m \ln^c(m/\epsilon))$ \cite{Cleve:1999vn}, so the dependence on the basis choice is at most logarithmic. 

It would be nice to establish  similar results in the QFT context, but so far almost nothing can be said about the gate dependence. In most of the works on the complexity, the choice of gates was restricted  to gates quadratic in the fields, in order not to depart from  the space of Gaussian states. No fundamental reason underlies this choice except that it allows to do explicit computations, and it is not obvious how to make an alternative choice: in general it is difficult to find an algebraically closed set of gates that is larger than the set of quadratic operators but not as large as the full space 
 \cite{Bhattacharyya:2018bbv,Cotler:2018ufx}.
 But it turns out that such a set is provided by the magic of bosonisation. 

The reason for this is that the bosonisation map is very non-linear,  so Gaussian states in one description generically do not correspond to Gaussian states in the other, and 
similarly gates that are  oscillator bilinears in one description will map to more complicated gates in the other. Thus we can have a completely solvable example where it is possible to explore the effect of 
different choice of gate sets on the complexity. 

%

We have started to undertaken this exploration in the present paper. We are not yet in a position to make general statements about the effect of a change of gates, since we are limited to the 
states whose complexity can be computed using the available technology.  
We  have identified two such classes of states. 
In the first class, the states are bosonic-coherent and fermionic-gaussian. For the states in this class, we can compute the complexity analytically, and we find that it has a similar form in the two descriptions, see the results in \eqref{eq:twoshift} and \eqref{eq:multimodefermion}, however the fermionic complexity appears to have a cost that depends on the mode number of the bosonic oscillator that is excited. Thus, the difference of complexity can become arbitrarily large, even with a single mode. 

In the second class, the states are both bosonic-gaussian and fermionic-gaussian \footnote{The existence of such states has not been remarked before, to our knowledge.}. These states can be understood as ground states of a system with a free but inhomogeneous hamiltonian. 
They are parametrized by an arbitrary function; we considered some examples with the function having only one or only two Fourier components. We cannot find analytic results in this case but have to resort to numerics. The numerical results are given in Figs. \ref{Level-two-bigauss},   \ref{fig:Fermion-level-two-cutoff} and  \ref{fig:bigaussiantwo}, and show striking differences between the 
bosonic and the fermionic result. The bosonic complexity for these states is cutoff-independent, and smoothly dependent on the parameters corresponding to the Fourier modes,  whereas for fermions it grows like $\ln(\Lambda)$, and it is much less regular (it appears to be quasi-periodic in the simplest case).


The plan of the paper is the following: in Section \ref{2Dbosonisation} we recall the basic properties of the bosonisation equivalence and the correspondence between fermionic and bosonic states. In Section \ref{sec:bosongates} we compute the complexity of a class of  bosonic coherent states  in terms of bosonic gates by using the \textit{Fubini-Study} metric method. 
In Section \ref{sec:nielsenreview} we review the \textit{Nielsen} method applied on free fermonic and bosonic field theory, and in Section \ref{sec:fermigates} we use it to compute the complexity of bosonic coherent states using fermionic gates.  In Section \ref{sec:gaussequiv} we describe a class of states that are of gaussian type both in the fermionic and in the bosonic description, and we compare the results for the complexity computed in either description. In Section \ref{Conclusions} we present our conclusions. Some additional details of the computations are presented in the Appendices.

\section{2D bosonisation}\label{2Dbosonisation}
\subsection{Basic results}
In this section, we will review the basics of the 2D bosonisation formalisms, for free bosons and fermions, which we will use in the rest of the paper. We follow the presentation in \cite{reviewbosonisation}.  In two dimensions, the bosonisation can be proved exactly for a  system on a finite size interval $[-L/2,L/2]$. In such a system, the unbounded momentum $k$ satisfies
\begin{equation}\label{eq:kANDn}
k= {2\pi \over L}\left(n_k -  {\delta_b \over 2} \right),~~~~~n_k\in \mathbb{Z},~~~~~\delta_b\in[0,2)
\end{equation}
where $\delta_b$ depends on the periodicity condition of the fermionic fields, $0$ for complete periodicity and $1$ for anti-periodicity. If there are $M$ chiral fermions with periodic conditions $(\delta_b=0)$ in the system, an index $\eta$ could be used to denote different types of fermions, and the mode decomposition  for each fermion type is given as
\begin{equation}\label{eq:fermionmodes}
\psi_\eta (x) = \left( {2\pi\over L} \right)^{1/2} \sum_{n=-\infty}^\infty e^{-i {2\pi n\over L} x}c_{n\eta},~~~~~c_{n\eta} = (2\pi L)^{-1/2} \int_{-L/2}^{L/2} dx e^{i{2\pi n\over L}x} \psi_\eta (x) \,,
\end{equation}
where $\eta = 1,2,\dots,M$, can be \textit{spin, handedness, flavor} etc. The bosonic chiral fields are given by the mode decomposition 
\begin{equation}\label{eq:bosonmodes}
\phi_\eta (x) = - \sum_{n>0} {1\over \sqrt{n}} (e^{-i {2\pi n\over L}x} b_{n\eta} + e^{i{2\pi n\over L}x} b_{n\eta}^\dagger)e^{-a{\pi n\over L}} \,,
\end{equation}
where the zero mode is omitted.  Notice that only $n>0$ modes are included, because of the chirality. 
The $a$ appearing in the last factor is a regularisation parameter that should be sent to zero when computing physical quantities.\footnote{Notice that the bosonisation is exact only when all the modes are included. In practice we often need to introduce a UV cutoff on the mode number; we still expect that we can match quantities that are cutoff-independent.} 

By construction, there is an operator identity at the level of annihilation and creation operators between fermions and bosons,
\begin{equation}\label{eq:Obosonfermion}
b_{n\eta} = {-i \over \sqrt{n}}\sum_{l= -\infty}^ \infty c^\dagger _{l-n ~\eta}c_{l \eta},~~~~~ b_{n\eta}^\dagger = {i \over \sqrt{n}}\sum_{l= -\infty}^ \infty c^\dagger _{l+n ~\eta}c_{l \eta}
\end{equation} 
from where we see that the bosonic operators are always an infinite sum of fermionic operators of quadratic type containing both the creation and annihilation ones. 
We can take \eqref{eq:Obosonfermion} as a definition of the bosonic modes; 
the proof of bosonisation amounts to showing that, with this definition, the  bosonic commutation relations are satisfied if the fermionic ones are: 
\begin{equation}
	\{c_{l \eta}, c_{\tilde{l} \tilde{\eta}}^\dagger \} = \delta_{\eta \tilde{\eta}}\delta_{l \tilde{l}}= [b_{l \eta},b_{\tilde{l} \tilde{\eta}}^\dagger ]. 
\end{equation}
The bosonisation formula can also be stated in terms of the local fields: 
\begin{equation}\label{eq:bosonfield}
i \partial_x \phi_\eta (x) = \normord{\psi^\dagger_\eta(x) \psi_\eta(x)} \,,
\end{equation}
with the colon denoting normal ordering. The inverse formula is more complicated: 
\begin{equation}\label{eq:fieldidentity}
	\psi_\eta (x)= F_\eta \, a^{-1/2} e ^{-i {2\pi\over L}(\hat{N}_\eta - {\delta_b \over 2})x} \normord{e^{-i\phi_\eta(x)}},
\end{equation}
where $\hat{N}_\eta = \sum_l :c^\dagger_{l~ \eta} c_{l~ \eta}:$ is the fermionic number operator with respect to the $\eta$th spiecies. 
From \eqref{eq:Obosonfermion}, it is easy to see that the bosonic operators commute with the fermionic number operator, \ie
\begin{equation}
[b_{n~\eta},\hat{N_\eta}]=0,~~~~~ [b_{n~\eta}^\dagger,\hat{N_\eta}]=0 \,,
\end{equation}
so only fermionic operators that don't change the fermion number can strictly speaking be bosonised. The so-called Klein factor 
 $F_\eta$ in \eqref{eq:fieldidentity} has the role of compensating the mismatch in fermion number (this factor is often omitted in many presentations of bosonisation). However we will not need 
 its explicit expression. 

The commutation relation between bosonic and fermionic modes is 
\begin{align}
[b_l, c_{\tilde{l}}] = {i\over \sqrt{l}}c_{\tilde{l}+l},~~~~~ [b^\dagger_l, c^\dagger_{\tilde{l}}] = {i\over \sqrt{l}}c^\dagger_{\tilde{l}+l},\\
[b_l, c^\dagger_{\tilde{l}}] = {-i\over \sqrt{l}}c^\dagger_{\tilde{l}-l},~~~~~ [b^\dagger_l, c_{\tilde{l}}] = {-i\over \sqrt{l}}c_{\tilde{l}-l}.
\end{align}
Intuitively we can think that $b_l^\dagger$ creates, and $b_l$ annihilates, a particle-hole pair of total momentum $l$. 

In the following, we will consider the simplest case with only two chiral fermions and one chiral bosons, thus the species index will be omitted.

\subsection{Fermionic Fock space } 
In the fermionic Fock space $\mathcal{F}$, a unique vacuum $\ket{0}$ is defined in terms of the fermionic modes,
\begin{align}\label{eq:fermivacuum}
c_n \ket{0} = 0,~~~ n>0;~~~~~c_n^\dagger \ket{0} = 0,~~~ n\le 0.
\end{align}
In the bosonised picture, the Fock space can be reorganised as a direct sum of all the Hilbert space with fixed fermionic particle number, i.e., 
\begin{equation}\label{eq:fockspace}
\mathcal{F} = \oplus_{ N}\mathcal{H}_{N}.
\end{equation}
Each $\mathcal{H}_{N}$ with fixed fermion number can be regarded as the bosonic Hilbert space since the bosonic operators commute with the fermionic number operator as mentioned before. The condition for a bosonic ground state  is to be annihilated by all the bosonic annihilation operators 
\begin{equation}\label{eq:bosonicground}
b_n \ket{\mathcal{G}_B}=0,~~~~~n>0.
\end{equation}
This condition uniquely defines a state in each module $\mathcal{H}_N$; then the ground state $\ket{\mathcal{G}_B}$ in the $N$-particle module is denoted as
\begin{align}\label{eq:bosonicGS}
\ket{\mathcal{G}_B^N} = \left\{ 
\begin{array}{rr}
c_{N}^\dagger c_{N-1}^\dagger\cdots c_{1}^\dagger \ket{0} ,& N>0\\
\ket{0}, &N=0\\
c_{N+1} c_{N}\cdots c_{0}\ket{0}, & N<0\\
\end{array}\right..
\end{align}

\begin{figure}
	\centering
	\centering \includegraphics[scale=0.5]{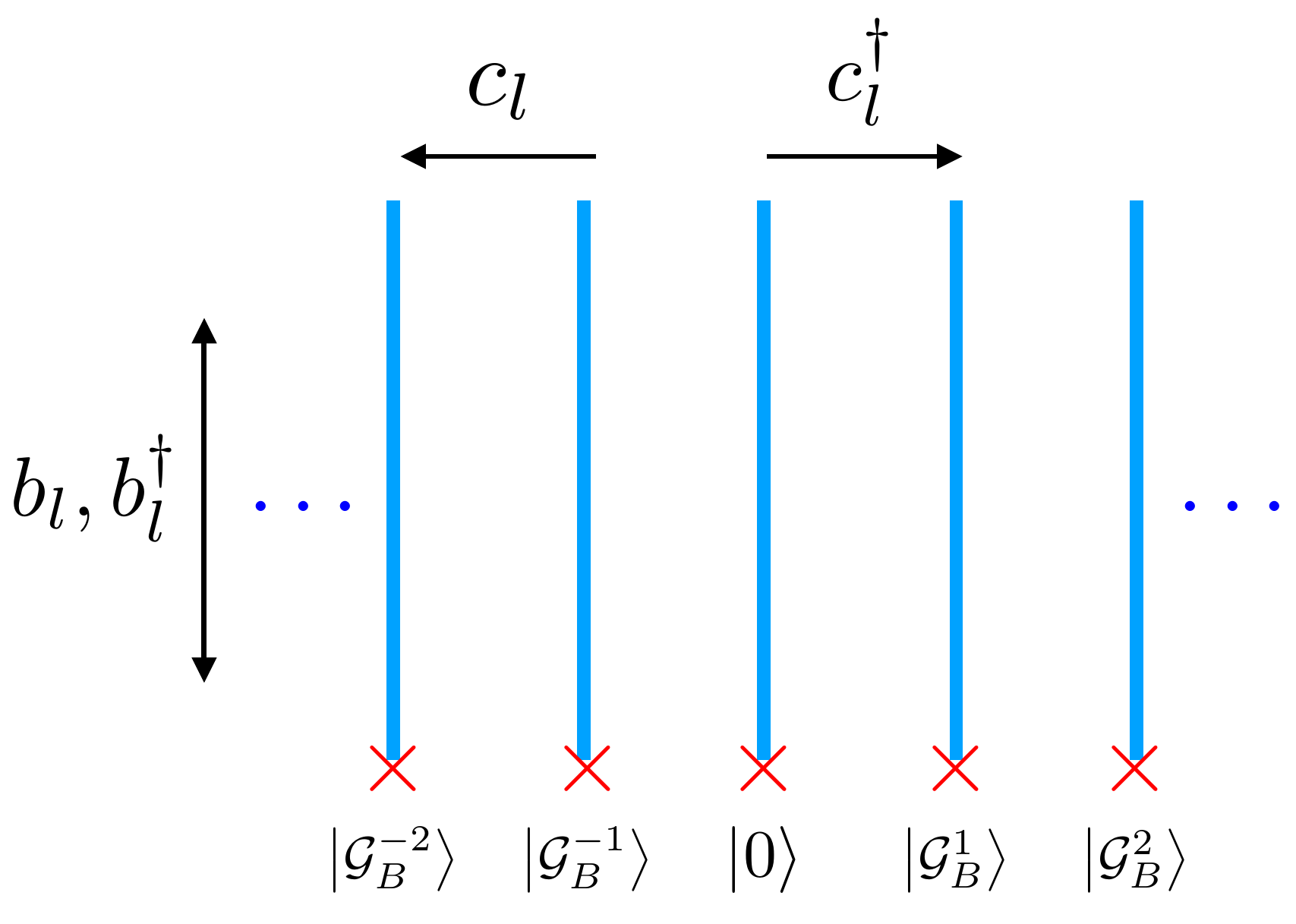}
	\caption{The base manifold $\mathcal{G}_B$ is made of all the bosonic ground states which are denoted as the red crosses.  The vertical blue lines are modules of the fermionic fock space where each represents a bosonic Hilbert space with fixed fermion number. The operators of bosonic type can only move vertically while the fermionic ones span the whole Fock space with a single fermionic operator moving horizontally.} 
	\label{fig:fockspace}
\end{figure}

Figure  \ref{fig:fockspace} gives a depiction of the Fock space as a bundle, with the base given by the ground states and the fiber by the bosonic Hilbert space. The bosonic operators act inside a 
single fiber, whereas fermionic operator can move between different fibers. 

%


\subsection{Correspondence between states}\label{sec:BFbridge}

Our main purpose in this note is to compare the circuit complexity of states in the bosonic and fermionic descriptions. We cannot consider the most general state, since the techniques developed so far 
only allow to deal with Gaussian states or coherent states. Our first task is to determine the correspondence between Gaussian/coherent states in both descriptions. 

It is clear that, since the bosonisation relations \eqref{eq:bosonfield}-\eqref{eq:fieldidentity} are non-linear, Gaussian states in one description will not correspond in general to Gaussian states in the other. 
It is shown in \cite{Hackl:2018ptj} that all Gaussian states can be obtained from the fermionic vacuum \eqref{eq:fermivacuum} by the action of a unitary operator of the form 
\begin{equation}
\mathcal{O}_F = e^{A^{lm}c_lc_m + B^{lm}c_lc_m^\dagger + D^{lm}c_l^\dagger c_m^\dagger} \,.
\end{equation}
Equivalently, each such states is a  vacuum under a set of modes related to the original ones by a Bogoliubov transformation
\begin{equation} 
\tilde c_l = {\cal A}_{lm} c_m + {\cal B}_{l m } c_m^\dagger \,. 
\end{equation}
It is easy to see then that states of the form 
\begin{equation}
c_{n_1}^\dagger c_{n_2}^\dagger \cdots c_{n_k}^\dagger c_{m_1} c_{m_2}\cdots c_{m_q}\ket{0} 
\end{equation}
with $n_1>n_2>\cdots>n_k>0\ge m_1 > m_2>\cdots >m_q$, are Gaussian states; these include in particular the bosonic ground states of  \eqref{eq:bosonicGS}. 

Another class of fermionic Gaussian states is given by 
\begin{equation}
\ket{\{\alpha_k,n_k\},N} = e^{\sum_k \alpha_k b_{n_k}^\dagger -\alpha_k^* b_{n_k}}\ket{\mathcal{G}_B^N},
\end{equation}
obtained from the ground state with the action of the {\it displacement operator} which is a fermion bilinear (see \eqref{eq:Obosonfermion}). These are bosonic coherent states, therefore not Gaussian but still tractable. 
In the next sections we will analyze the complexity of the last class of states, first from the bosonic and then from the fermionic perspective. 

%


\section{\textit{Fubini-Study metric} method for Bosonic Coherent states}\label{sec:bosongates}

The complexity of coherent states of a set of harmonic oscillators was considered in \cite{Guo:2018kzl} using both the Nielsen approach and the Fubini-Study approach. In order to use the Nielsen's approach with bosonic gates, we would need to embed the coherent states in a larger space acted on by  $GL(n)$, therefore in this section we will use the FS method for simplicity.  
The results of \cite{Guo:2018kzl} show that the FS complexity is 
very closely related to the Nielsen complexity with the cost function given by the invariant metric on the group. See also \cite{Yang:2017nfn} for a different approach based on Finsler geometry. 

In the FS approach, the complexity $\mathcal{C}(| \psi_1\rangle, |\psi_2\rangle )$ is computed as the geodesic distance between two states $| \psi_1\rangle$ and $| \psi_2\rangle$, with respect to the Fubini-Study metric on the projective Hilbert space $\cal{H}/ \mathbb{C}^*$. 
The metric can be described as follows: given a parametrised path  $\ket{\psi(\sigma)}$ on the manifold of normalized states, the line element, \ie \, the length of the tangent vector to the path, 
is 
\begin{equation}\label{eq:FSline}
	ds = d\sigma \sqrt { \braket{ \der_\sigma \psi(\sigma) | \der_\sigma \psi(\sigma) }
	 - |\braket{\psi(\sigma) | \der_\sigma \psi(\sigma) }|^2} \,.
\end{equation}
Equivalently, for a family of states $\psi(\lambda)$ parametrized by coordinates $\lambda^i$, the FS metric is $g_{ij} d\lambda^i d\lambda^j$, with 
\begin{equation}\label{eq:FSfamily}
	g_{i j}  = \braket{ \der_{(i} \psi | \der_{j)} \psi } - \braket{\der_i \psi | \psi} \braket{\psi| \der_j \psi} \,.
\end{equation}


Let us consider the coherent states as described in the previous section. For simplicity we start with the case of a displacement operator acting only on one mode of the bosons. 
We consider the one-complex-parameter family of states 
\begin{equation}\label{eq:onemode}
	\ket{\psi(\alpha)} =\ket{\alpha,n,N} = U_n(\alpha)\ket{\mathcal{G}_B^N} \,,
\end{equation}
where the displacement operators 
\begin{align}
U_n(\alpha)&= e^{\alpha b^\dagger_n - \alpha^{\ast} b_n} = e^{\alpha b^\dagger_n}e^{-\alpha^\ast b_n}e^{-{1\over 2} \alpha \alpha^\ast},\\
U^\dagger_n(\alpha) &= e^{-\alpha b^\dagger_n + \alpha^{\ast} b_n} = e^{-\alpha b^\dagger_n}e^{\alpha^\ast b_n}e^{-{1\over 2} \alpha \alpha^\ast} 
\end{align}
satisfy  
\begin{align}
	\der_\alpha U_n(\alpha) =  b_n^\dagger \, U_n(\alpha) - {1 \over 2}  \alpha^\ast \, U_n(\alpha), & \quad
	\der_{\alpha^\ast} U_n(\alpha) =  -  U_n(\alpha) b_n - {1 \over 2} \alpha \, U_n(\alpha), \\
	\der_\alpha U^\dagger_n(\alpha) = -  b^\dagger_n U^\dagger_n(\alpha) - {1 \over 2}  \alpha^\ast \, U^\dagger_n(\alpha) , & \quad
	\der_{\alpha^\ast} U^\dagger_n(\alpha) =  U^\dagger_n(\alpha) b_n - {1 \over 2} \alpha \, U^\dagger_n(\alpha) \,.
\end{align}
It is useful to notice the  commutation relations
\begin{align}
[b_n,U_l(\alpha)] = \delta_{ln} \, \alpha \, U_n(\alpha),~~~~~ [U^\dagger_l(\alpha), b^\dagger_n] = \delta_{ln}\, \alpha^\ast \, U^\dagger_n(\alpha) \,, 
\end{align}
from which it follows that 
\begin{equation} 
\langle U_n^\dagger (\alpha) b_n U_n(\alpha) \rangle = \alpha \,, \quad  \langle U_n^\dagger (\alpha) b_n^\dagger U_n(\alpha) \rangle = \alpha^\ast \,, 
\end{equation}
where $\langle \ldots \rangle$ denotes the vev in the state $\ket{\mathcal{G}_B^N}$. 
The FS metric components can be computed: 
\begin{align}
 g_{\alpha \alpha} = & \langle \der_\alpha U_n^\dagger(\alpha) \,  \der_\alpha U_n(\alpha) \rangle - \langle \der_\alpha U_n^\dagger(\alpha) \,  U_n(\alpha) \rangle \langle  U_n^\dagger(\alpha) \der_\alpha U_n(\alpha) \rangle  = 0 \,, \\
 g_{\alpha^\ast \alpha} = & \langle \der_{\alpha^\ast} U_n^\dagger(\alpha) \,  \der_\alpha U_n(\alpha) \rangle - \langle \der_{\alpha^\ast} U_n^\dagger(\alpha) U_n(\alpha) \rangle \langle  U_n^\dagger(\alpha) \der_\alpha U_n(\alpha) \rangle  = 1 \,. 
 \end{align}
We obtain then that the FS metric is a flat K\"ahler metric\begin{equation}
ds^2 = d\alpha \, d\alpha^\ast,
\end{equation}
as  expected, since  it is known that in quantum mechanics coherent states form a two-dimensional \textit{K\"ahler manifold} which can be parametrized by classical phase space variables
 \cite{Bengtsson:2006:GQS:1211332}.

For our purposes it is important to notice that the result does not depend on either the fermion number $N$, or the bosonic oscillator number $n$. 
We can easily understand the $N$-independence since all the bosonic Hilbert spaces $\cal{H}_N$ are equivalent to each other. The independence on the mode number indicates that the FS metric 
corresponds to a cost function that assigns the same cost to all displacement operators $U_n$. This is also natural since the FS metric is derived from the scalar product in the Hilbert space, which has no information about the energy levels apart from the fact that they are orthogonal to each other. 
It is trivial to compute the geodesic length in the flat metric: 
\begin{equation}\label{eq:bosonresultOM}
	\mathcal{C}_{FS} (\ket{\mathcal{G}_B^N} , \ket{\psi(\alpha) }) = |\alpha| \,.
\end{equation}
This agrees with the result of  \cite{Yang:2017nfn}; the result of  \cite{Guo:2018kzl} is   $\mathcal{C}_{FS}(\alpha) = \textrm{arccosh} (1+\frac{|\alpha|^2}{2})$, which agrees with \eqref{eq:bosonresultOM} for small $\alpha$, while for larger $\alpha$ it is smaller,  which means that one can find shorter geodesics if the coherent states are embedded in the larger space mentioned at the beginning of this section. 



The general case of coherent states contains displacement operators acting on several modes: 
\begin{equation}\label{eq:coherentexample}
\ket{\psi(\alpha)}=\prod_{i=1} ^k U_{n_i}(\alpha_i)\ket{\mathcal{G}_B^N} \,,
\end{equation}
with $n_i \neq n_k$ for $i \neq k$. 
In order to compute the metric it is enough to consider the case of two modes: 
\begin{equation}
\ket{\psi(\alpha,\beta)} =\ket{\alpha,\beta;m,n;N} =U(\alpha,\beta) \ket{\mathcal{G}_B^N} \,, \quad U(\alpha,\beta)= U_m(\alpha)U_n(\beta) \,.
\end{equation}
The off-diagonal metric components are 
\begin{align} 
g_{\alpha\beta} &= \frac{1}{2} \langle \der_\alpha U^\dagger \der_\beta U \rangle + \frac{1}{2} \langle \der_\beta U^\dagger \der_\alpha U \rangle - \langle \der_\alpha U^\dagger U \rangle \langle U^\dagger \der_\beta U \rangle = 0 \,, \\
 g_{\alpha^\ast\beta} &= \langle \der_\alpha^\ast U^\dagger \der_\beta U \rangle  - \langle \der_\alpha^\ast U^\dagger U \rangle \langle U^\dagger \der_\beta U \rangle = 0 \,.
 \end{align}
It is easy to see that since the displacement operators of different modes commute, this applies for any number of excited modes. Therefore the metric is a direct product of the metrics for the individual modes: 
\begin{align}
	ds^2 = \sum_i d\alpha_i d\alpha_i^\ast 
\end{align}
and the complexity of a general coherent state is 
\begin{align}\label{eq:twoshift}
	\mC_{FS}(\ket{\mathcal{G}_B^N} ,  \prod_i ^k U_{n_i}(\alpha_i)\ket{\mathcal{G}_B^N}) = \sqrt{\sum_{i=1}^{k}|\alpha_i|^2} \,.
\end{align}

\section{\textit{Nielsen} method for gaussian states}\label{sec:nielsenreview}
The \textit{Nielsen} method can be understood as a way to geometrize the notion of complexity. In this section, we will review the Nielsen method and its application to fermionic and bosonic  Gaussian states, along the lines developed  in \cite{Hackl:2018ptj,Jefferson:2017sdb,Chapman:2018hou}. 

In Nielsen's method, the circuit is replaced by a continuous path of unitary operators 
\begin{equation} 
U(s) = \mathcal{P}\textrm{exp} \left( -i \int_0^s ds' \, Y^I(s') \mathcal{O}_I \right) \,,
\end{equation} 
such that $U(s=1)$ gives the desired unitary operator (i.e., implements the circuit we want), and $\mathcal{O}_I$ is a basis of operators that can be used to build the circuit (we can think of them as the generators of the infinitesimal gates $e^{i \epsilon \mathcal{O}_I}$). 
One can think that the final state is being prepared by means of an evolution in a fictitious time $s$ with a time-dependent Hamiltonian 
$H = Y^I(s) \mathcal{O}_I$. 
The complexity is then determined by the choice of a suitable {\it cost functional}  
\begin{equation} 
\mathcal{C} [U(s) ]  = \int_0^1 F \left(U(s), Y(s) \right) \,.
\end{equation}
The minimization of this functional will determine the form of the functions $Y^I(s)$ and thus the optimal circuit. 
With an appropriate choice, the cost functional determines a distance on the space of unitary operators, and the problem of minimizing the complexity of a circuit is mapped to the 
problem of finding a geodesic path on a Riemannian manifold. The cost functional can for instance have the form 
\begin{equation} \label{riemann-cost}
F_2 \left(U(s), Y(s) \right)  = \sqrt{G_{IJ}(U(s)) Y^I(s) Y^J(s)} \,.
\end{equation} 
Often the metric $G_{IJ}(U)$ is choosen to be a right-invariant metric, but not necessarily the canonical biinvariant metric defined by $\langle A, B \rangle = tr(AB)$ for $A,B$ self-adjoint. 
In fact, the possibility of choosing different metrics can be thought of as introducing penalty factors for moving along certain directions, corresponding to gates that may be more difficult 
to implement (for instance, in a system of qubits, one wants to penalize gates that act on many qubits simultaneously). 
A more general choice of the cost functional would lead to a Finsler geometry, where there is notion of a distance but not induced by a metric on the tangent space \cite{Jefferson:2017sdb}. 
For instance one can consider the family of cost functionals dependent on a parameter $\kappa$: 
\begin{equation}\label{kappa-cost}
F_\kappa  \left(U(s), Y(s) \right) = \sum_I |Y^I |^\kappa \,.
\end{equation}
Sometimes the ambiguity is reduced: the functional \eqref{kappa-cost} with  $\kappa=1$ is parametrization-independent. 
Another cost functional, which we will use in this paper, is  basis-independent and makes use of the {\it Schatten p-norm}:\footnote{Despite a certain resemblance, the Schatten $p$-norm is not directly related to the $\kappa$ cost function \eqref{kappa-cost}, although one might speculate that the two give the same result for a particular choice of basis.}
\begin{align}
F_p  \left(U(s), Y(s) \right) & = || Y^I(s) \mathcal{O}_I ||_p \,, \\
|| A ||_p \equiv & \left(\textrm{tr} (A^\dagger A)^{p \over 2}\right)^{1\over p} \,.
\end{align}

\subsection{Application on gaussian states in free fermionic field theory}\label{sec:FGreview}
In Nielsen's approach, the metric is defined not on the space of states, but on the manifold of a group of operators that act on the states. For fermions, the relevant group turns out to be 
the orthogonal group \cite{Hackl:2018ptj}. To see this, it is convenient to use  the Majorana basis
\be\label{eq:Majoranabasis}
\xi^a= \{q_{+\infty},\dots,q_{1},q_0,q_{-1},\dots,q_{-\infty},p_{+\infty},\dots, p_{1},p_0,p_{-1}\dots p_{-\infty}\}
\ee
which is related to the annihilation and creation modes  in the following way
\begin{align}
	c_l^\dagger &= {1\over \sqrt{2}}(q_l -i p_{l}),~~ c_l = {1\over \sqrt{2}}(q_l + i p_{l}).
\end{align}
In the Majorana basis the anticommutation relations read $\{\xi^a, \xi^b\} = \delta^{ab}$.  

A general Gaussian state $\ket{\psi}$ is completely characterized by the two-point function, or covariance matrix, 
\begin{equation}\label{eq:covardef}
\bra{\psi}\xi^a\xi^b\ket{\psi} = \frac{1}{2}(G^{ab}+i\Omega^{ab}_{\psi})
\end{equation}
which can be decomposed in the symmetric and antisymmetric part, respectively $G^{ab}$ and $\Omega^{ab}_\psi$. The symmetric part is determined by the anticommutation relation, therefore it is state-independent, 
while the antisymmetric part  encodes the state. 
In terms of the covariance matrix, the fermionic vacuum state $\ket{0}$ defined in \eqref{eq:fermivacuum} would be expressed as,
\begin{align} \label{G-Omega-C}
G^{ab}=\delta^{ab},~~ \Omega_{0}^{ab} =  \begin{pmatrix}
	0 & C\\
	-C & 0
	\end{pmatrix}~ \text{with} ~ C = 	
	\begin{pmatrix}
	\mathbb{1} & 0\\
	0 & -\mathbb{1}
	\end{pmatrix}. 
\end{align}
As discussed in sec. \ref{sec:BFbridge} the Gaussian states are obtained by Bogoliubov transformations; in the Majorana basis, these are linear transformations $\tilde \xi = M \xi$ that leave the 
anticommutation relation invariant, therefore they are orthogonal transformations. If we put a UV cutoff so that there are $2N$ modes, $M \in SO(2N)$.\footnote{We consider for simplicity only the component connected with the identity. It turns out that this is the subgroup that does not change the parity $(-1)^F$ of the fermion number of the state.}  However there is a subgroup that acts trivially on the state, which only change the state by a phase. The covariance matrix transforms as 
\begin{equation} 
\tilde \Omega = M \Omega M^T \,. 
\end{equation} 
The covariance matrix of the vacuum can be considered as a symplectic form, from this we see that the orthogonal matrices that leave $\Omega$ invariant  should be symplectic as well, \ie, they should be unitary. The 
manifold of Gaussian states is then $SO(2N)/U(N)$. 
A convenient way to parametrize this coset is by using the relative covariance matrix 
\begin{equation}\label{relative-covariance}
\Delta(M) = \tilde \Omega \, \Omega_0^{-1} = M \Omega_0 M^T \Omega_0^{-1}. 
\end{equation} 
Conversely, if we are given the reference and target states $\Omega_0$ and $\tilde \Omega$, we can recover the transformation matrix up to a unitary transformation. The polar decomposition of 
orthogonal matrices allows us to write $M=u\hat{M}$ with $u \in U(N)$ and $\hat{M} $ antisymmetric. In this manner, $u$ and $\hat{M}$ are uniquely defined by the polar decomposition. 

At the level of Lie algebra, the splitting $so(2N) = u(N) \oplus \textrm{asym(N)}$ is an orthogonal decomposition with respect to the Killing metric of $so(2N)$. We expect then that the shortest path in the coset space will be obtained by moving only along the second subspace. Indeed it is shown in \cite{Hackl:2018ptj} that the geodesic connecting $\Omega_0$ and $\tilde\Omega$ is given by a straight curve $\gamma(s) = e^{s A}$ with $s\in[0,1]$ which has a constant direction $A\in\textrm{asym(N)}$.  Finally the geodesic length is given by the norm of $A = \frac{1}{2}\ln\Delta$, which is the inner product with itself $||A|| = \sqrt{\left<A,A\right>}$ using the embedded metric on the Lie manifold $SO(2N)$. This definition of norm is basis independent and will coincide with the \textit{Schatten} $p=2$  norm. The \textit{Schatten} $p-$norm for a general matrix $T$ is defined as
\begin{equation}
||T||_p = \left( \sum_{n\ge1} s^p _n(T) \right)^{1/p}
\end{equation}
with $s_n(T)$ being the singular values of the $n\times n$ matrix $T$, \ie, the eigenvalues of the matrix $\sqrt{T^\dagger T}$. Due to the decreasing monotonicity of the \textit{Schatten} $p-$norm, an interesting case is the $p=1$ norm, defined as
\begin{equation}
||T||_p = \text{Tr}(\sqrt{T^\dagger T} )= \sum_{n\ge1} s_n(T) 
\end{equation}
which will impose an upper bound to the $p=2$ norm. The $p=1$ norm has been considered before for quantum information purposes, for example it has been found that it is the only one among the $p$-norms to provide a consistent measure for quantum correlations, called quantum discord \cite{PhysRevA.87.064101}.

%
To summarize, the complexity of fermionic Gaussian states, defined as the geodesic length with respect to the Killing metric on the orthogonal group: 
\begin{align}
\mathcal{C}_f(\ket{0}, \ket{\psi})  \equiv ||A|| &= ||A||_{p=2} =  {1\over 2} \sqrt{\text{Tr}|(i\ln \Delta)^2|} = \frac{1}{2} \sqrt{\sum_r (i\ln \lambda_r)^2}\label{eq:complexityp2}
\\
&\le ||A||_{p=1} = \frac{1}{2} \text{Tr} |i\ln \Delta| = \frac{1}{2}\sum_r |i\ln \lambda_r|.\label{eq:complexityp1}
\end{align}
In the above formula $\lambda_i$ are the eigenvalues of $\Delta$ which come in pairs $e^{\pm i \theta}$ since $\Delta\in SO(2N)$. Although the $p=1$ norm loses its geometric meaning in the current case, it is interesting because in some cases discussed in the later sections, it poses an analytical bound on the Gaussian state complexity.

\subsection{Translation to gaussian states in free bosonic field theory}
In the case of free bosonic theory with $N$ degrees of freedom, the gates group corresponding to Gaussian states is $Sp(2N)$. Contrary to the fermionic case,  the role of covariance matrices $G$ and $\Omega$ defined in \eqref{eq:covardef} is exchanged, in the way that $G$ represents the state and $\Omega$ encodes the algebraic relation. The main objective is still to find the relative covariance matrix relating the reference state and the target state, given as
\begin{equation} \label{relative-covariance-bos}
\Delta =\tilde{G} G_0 =  M G_0 M^T G_0^{-1}
\end{equation}
where $M\in Sp(2N)$ encodes the basis transformation. Similar to the fermionic case, $\Delta$ is an element in the coset space $Sp(2N)/U(N)$ and the geodesic connecting $\tilde{G}$ and $G_0$ will be again $\gamma(s) = e^{sA}$ with however $A\in \text{sym}(N)$ being an element in the symmetric algebra. In the same manner, the geodesic length is given by the norm of $A= \frac{1}{2}\ln\Delta$, hence the complexity
\begin{align}
\mathcal{C}_b(\ket{0}, \ket{\psi})  \equiv ||A|| &= ||A||_{p=2} =  {1\over 2} \sqrt{\text{Tr}|(\ln \Delta)^2|} = \frac{1}{2} \sqrt{\sum_r (\ln \lambda_r)^2}\label{eq:bcomplexityp2}
\\
&\le ||A||_{p=1} = \frac{1}{2} \text{Tr} |\ln \Delta| = \frac{1}{2}\sum_r |\ln \lambda_r|.\label{eq:bcomplexityp1}
\end{align}
Compared to the eq. \eqref{eq:complexityp2}, there is an ``$i$'' difference, which is due to the fact that the eigenvalues of a real symmetric symplectic matrix are in pairs of  $e^{\pm r}$ with $r\in\mathbb{R}$. If the covariance matrix $G$ for the reference state is the identity matrix, the covariance matrix will simply be $\Delta = M  M^T$, which we shall use in section \ref{sec:gaussequiv}. Detailed studies of the free bosonic complexity can be found in \cite{Chapman:2018hou,Hackl:2018ptj}.

\section{Application of \textit{Nielsen} method on bosonic coherent states}\label{sec:fermigates}
In this section, we apply the method introduced in the last section \ref{sec:FGreview} on the bosonic coherent states with instead fermionic gates.
\subsection{Complexity between bosonic ground states}
As a first simple application of the formalism, we can compute the complexity of the bosonic ground states $\mathcal{G}_B^N$  defined in \eqref{eq:bosonicGS}, which constitute a subset of the 
fermionic gaussian states, as already noticed. These states are labeled by the fermion number $N$. The Bogoliubov transformation that takes from  $\ket{\mathcal{G}_B^0}$ to  $\ket{\mathcal{G}_B^m}$ is 
(for $m$ positive)
\begin{align}
\tilde c_l = c_l^\dagger \,, ~~~~~ \tilde c_l^\dagger = c_l \,, ~~~~ 1 \leq l \leq m \,.
\end{align}
The corresponding matrix $M$ changes the sign of $p_l, 1 \leq l \leq m$ and it is in $SO(2N)$ only if $m$ is even. As noticed before, we can only find geodesic paths between states that have the same 
parity of fermion number. 

The covariance matrix $\Delta$ in this case is the identity, except for $2n$ diagonal entries that are equal to $-1$. Applying \eqref{eq:complexityp2} we find 
\begin{equation}\label{eq:GBcomplexity}
\mathcal{C}(\ket{\mathcal{G}_B^m} ,  \ket{\mathcal{G}_B^{m+2k}}) = 2 \pi  |k| \,.
\end{equation}
As observed before, the bosonic operators act vertically in the fibers, so there is no corresponding bosonic complexity in this class of states. 

\subsection{Bosonic coherent states with one excited mode} \label{sec:onemodeferm}
We consider next the coherent states analysed in section \ref{sec:bosongates}, but now in the fermionic description. We start again from the coherent states involving excitations of only one bosonic mode, namely the states $U_n(\alpha) \ket{\mathcal{G}_B^N}$ and consider within the states module of zero fermion number $N=0$, the analysis can be applied in the same manner to the other states module having a different fermion number with only a shift in the fermionic modes. 

As discussed in Section \ref{sec:BFbridge}, the unitary operator $U_n(\alpha)=e^{\alpha b_n^\dagger - \alpha^* b_n}$ is the exponential of a fermion bilinear, therefore the states are fermionic gaussian and the formalism developed in the first part of this section can be applied.  Using the Baker-Campbell-Hausdorff formula
\begin{equation}
e^X Y e^{-X} = Y+[X,Y] + {1 \over 2!}[X,[X,Y]] + {1\over 3!}[X,[X,[X,Y]]]+ \cdots,
\end{equation}
we can find the transformation of the oscillators in closed form, writing $\alpha = |\alpha | e^{i \theta}$ we have
\begin{align}\label{eq:basisrel1}
\tilde{c}_q(n)&=U_n(\alpha) c_q U_n^\dagger(\alpha)= \sum_{l,m\ge 0} \frac{(\alpha^*)^l \alpha^m}{m! l!} \left(\frac{-i}{\sqrt{n}}\right)^{l+m} c_{q+(l-m)n}\no\\
&=\sum_{r\ge 0} (-i)^r e^{-i r \theta} J_r\left({2 |\alpha|\over \sqrt{n} } \right)c_{q+nr} + \sum_{r< 0} i^r e^{-i r \theta} J_{-r}\left({2 |\alpha|\over \sqrt{n} } \right)c_{q+nr},\\
\tilde{c}_q^\dagger(n) &= U_n(\alpha) c_q^\dagger U_n^\dagger(\alpha)= \sum_{l,m\ge 0} \frac{(\alpha^*)^l \alpha^m}{m! l!} \left(\frac{i}{\sqrt{n}}\right)^{l+m} c_{q -(l-m)n}^\dagger\no\\
&=\sum_{r\ge 0} i^r e^{i r \theta} J_r\left({2 |\alpha|\over \sqrt{n} } \right)c_{q+nr}^\dagger + \sum_{r< 0} (-i)^r e^{i r \theta} J_{-r}\left({2 |\alpha|\over \sqrt{n} } \right)c_{q+nr}^\dagger, \label{eq:basisrel2}
\end{align}
where $J_r\left({2 |\alpha|\over \sqrt{n} } \right)$ is the Bessel function of the first kind and $n$ is the excitation mode of boson. Notice that the mixing occurs only within fermionic modes that differ by a multiple of $n$. In the Majorana basis the transformation reads 
\begin{align}\label{eq:ONrelation1}
\tilde{q}_l &= \sum_r J_{|r|}\left({2 |\alpha|\over \sqrt{n} } \right) \cos\left( {\pi \over 2}|r| + r \theta \right) q_{l+nr} + \sum_r J_{|r|}\left({2 |\alpha|\over \sqrt{n} } \right) \sin\left( {\pi \over 2}|r| + r \theta \right) p_{l+nr},\\
\tilde{p}_l &= \sum_r J_{|r|}\left({2 |\alpha|\over \sqrt{n} } \right) \cos\left( {\pi \over 2}|r| + r \theta \right) p_{l+nr} - \sum_r J_{|r|}\left({2 |\alpha|\over \sqrt{n} } \right) \sin\left( {\pi \over 2}|r| + r \theta \right) q_{l+nr} \,,\label{eq:ONrelation2}
\end{align}
and the corresponding orthogonal matrix has the form 
\begin{align} \label{Mferm-one}
M(n)= \left(\begin{array}{cc}
A(n) & B(n)\\
-B(n) & A(n)
\end{array}\right)
\end{align}
which is indeed orthogonal as we show in appendix \ref{app:OrthM} for the case that $\theta=0$.
Recall that the fermionic Gaussian state $\ket{\psi}$ is given in terms of the transformed oscillators by 
\begin{align}\label{eq:refstate}
\tilde{c}_q(n)\ket{\psi} = 0,~~~q>0;~~~~~\tilde{c}_q^\dagger (n)\ket{\psi} =0 ,~~~q\le0.
\end{align}
We observe that the $\theta$-dependence in \eqref{eq:basisrel1},\eqref{eq:basisrel2} can be eliminated by the following field redefinition 
\begin{align}\label{eq:phaseabsorbtion}
 c_q \to e^{iq\theta/n} c_q \,,  ~~ ~ ~ \tilde{c}_q(n) \to e^{iq\theta/n}\tilde{c}_q(n) \,, 
\end{align}
which does not affect the state \eqref{eq:refstate}. It is a unitary transformation, so it does not affect the complexity, as discussed in section \ref{sec:FGreview}. This is also in agreement with the result for the bosonic complexity \eqref{eq:bosonresultOM} which is independent of the phase of $\alpha$. We find that the most convenient choice for the present calculation is to set $\theta = \pi/2$. 
In this case the off-diagonal block $B(n)$ of the transformation matrix $M(n)$ vanishes identically, and the non-vanishing entries of $A(n)$ are given by
\begin{align}\label{eq:seccase}
A(n)_{i(i+nj)} &=\left\{
  \begin{array}{lr}
     J_{|j|}\left({2 |\alpha|\over \sqrt{n} } \right) & ~~(j\le 0) \\
     (-1)^{j}J_{|j|}\left({2 |\alpha|\over \sqrt{n} } \right) & ~~( j>0)\\
  \end{array}
\right.,
\end{align}
The relative covariance matrix $\Delta(n)$ is made of two identical blocks
\begin{align}\label{eq:delta2}
	\Delta(n) =   \begin{pmatrix}
	A(n)C A^T(n)  C & 0\\
	0 & A(n)C A^T(n) C
	\end{pmatrix},
\end{align}
with $C$  given in \eqref{G-Omega-C}. 
We can observe that $A^T(n)$ and $A(n)$ are related by an orthogonal transformation
\begin{align}
A^T(n) = O^T A(n) O
\end{align}
with $O_{ij}=(-1)^i\delta_{ij}$, which is symmetric and also commutes with $C$. The diagonal block can be rewritten as $(A(n)OC)^2$, so the problem is reduced to the diagonalization of $A(n) O C$. 

We have not been able to obtain the eigenvalues analytically. Instead, we notice that the matrix elements  $A(n)_{ij}$ (and those of $A(n) OC$ as well, they just differ by some minus signs) 
are of the form of Bessel functions $J_\nu(z)$ with index $\nu = |i-j|$ and argument $z= \frac{2\alpha}{\sqrt{n}}$.  
Recalling the power series expansion 
\begin{equation}
J_\nu(z) = \left({z\over 2} \right)^\nu\sum_{k=0}^\infty \frac{\left(- {z^2\over 4} \right)^k}{k! \Gamma(\nu+k+1)},
\end{equation}
we see that for small $z$ the matrix is dominated by the diagonal elements, while off-diagonal ones are suppressed exponentially with the distance from the diagonal. 
This has the consequence that the eigenvalues can be computed numerically with good accuracy, and they are weakly dependent on the cutoff that we put on the length of the matrix considered. One has to keep in mind that each entry in eq. \eqref{eq:delta2} contains the full contributions of all the modes. 
 
\begin{figure}[h!]
	\centering
	\includegraphics[scale=0.5]{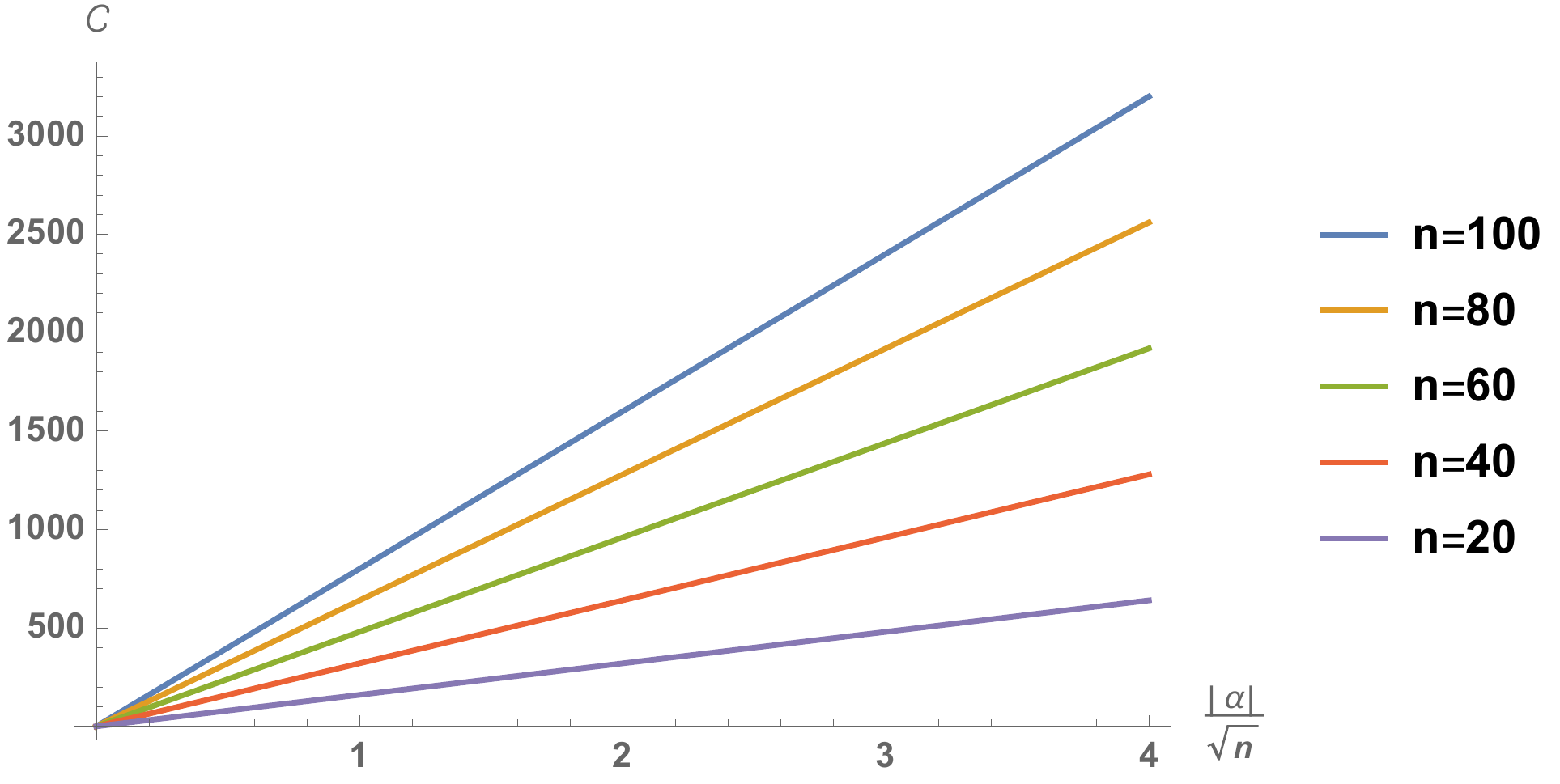}
	\caption{Complexity $\mathcal{C}_{p=1}$ over ${\alpha\over \sqrt{n} }$ is constant in $|\alpha|$,  for $n\in[20,100]$ with an interval of 20, where $n$ labels the bosonic excitation $b_n^\dagger$ in \eqref{eq:onemode}. The cutoff on the size of the matrices $A(n)$ and $B(n)$ is set equal to $10n$.}
	\label{fig:largeN}
\end{figure}

\paragraph{Result with $p=1$ norm}In Fig. \ref{fig:largeN} we plot the complexity with $p=1$ norm, as a function of $\alpha$, for different values of the number $n$, the excited bosonic mode. The cutoff is chosen to be $10 n$, although the result does not depend on it. We observe that the ratio  between the complexity $\mathcal{C}_{p=1}$ and ${ |\alpha|\over \sqrt{n} }$ which is the argument of the Bessel function, is constant in $\alpha$. 

 Therefore, we claim that the complexity between the bosonic ground state $\ket{\mathcal{G}_B^N}$ and the coherent state $\ket{\alpha,n}$ generated from it by using the \textit{Schatten} $p=1$ norm is  
\begin{equation}\label{eq:singlemodefermion}
	\mathcal{C}_{p=1}(\ket{\mathcal{G}_B^N}, U_n(\alpha)\ket{\mathcal{G}_B^N}) = 4\sqrt{n} |\alpha|
\end{equation}
where we have taken into account the two diagonal blocks and an overall factor $1/2$ in \eqref{eq:complexityp1}. Comparing to the result \eqref{eq:bosonresultOM} obtained in terms of bosonic gates, we see that the fermionic complexity has an extra dependence on the excitation mode $n$.

\begin{figure}[h!]
	\centering
	\includegraphics[scale=0.5]{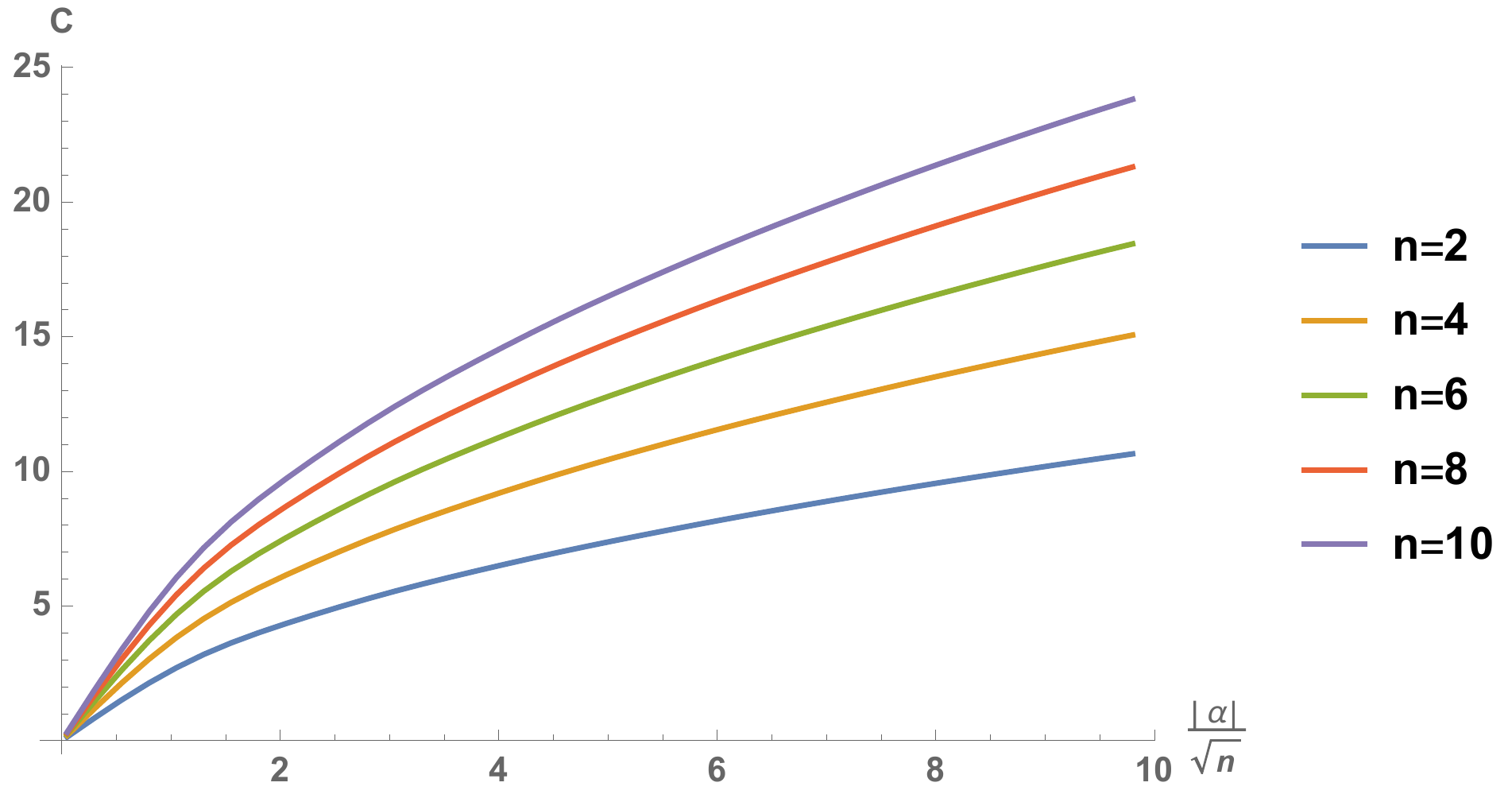}
	\caption{Complexity $\mathcal{C}_{p=2}$  increases when $|\alpha|$ increases for $n\in[2,10]$ with an interval of 2, where $n$ labels the bosonic excitation $b_n^\dagger$ in \eqref{eq:onemode}. The length of the matrices $A(n)$ and $B(n)$ is cut to be twenty times of the bosonic excitation $n$ for the plot.}
	\label{fig:cohp2}
\end{figure}

\paragraph{Result with $p=2$ norm}The \textit{Schatten} $p=2$ norm, as aforementioned, endows the complexity with the meaning of geodesic length on the gates' manifold. In Fig. \ref{fig:cohp2}, we plot the complexity similar to the previous case with different number of excitations $n$. We see here that the complexity is still increasing but not at all in a linear manner. 
The rate of the increase in each curve decreases with $|\alpha|$,  consistently with the monotonicity of the $p-$norms. 

Finally the complexity with the $p=2$ norm is bounded by $p=1$ norm result
\begin{equation}
\mathcal{C}(\ket{\mathcal{G}_B^N}\to U_n(\alpha)\ket{\mathcal{G}_B^N} ) = \mathcal{C}_{p=2}(\ket{\mathcal{G}_B^N}\to U_n(\alpha)\ket{\mathcal{G}_B^N}) \le 4\sqrt{n} |\alpha|.
\end{equation}

\subsection{Complexity for bosonic coherent states with shifts in more modes}
\begin{figure}[h!]
	\centering
	\begin{subfigure}{0.48\linewidth}
 	\includegraphics[width=\linewidth]{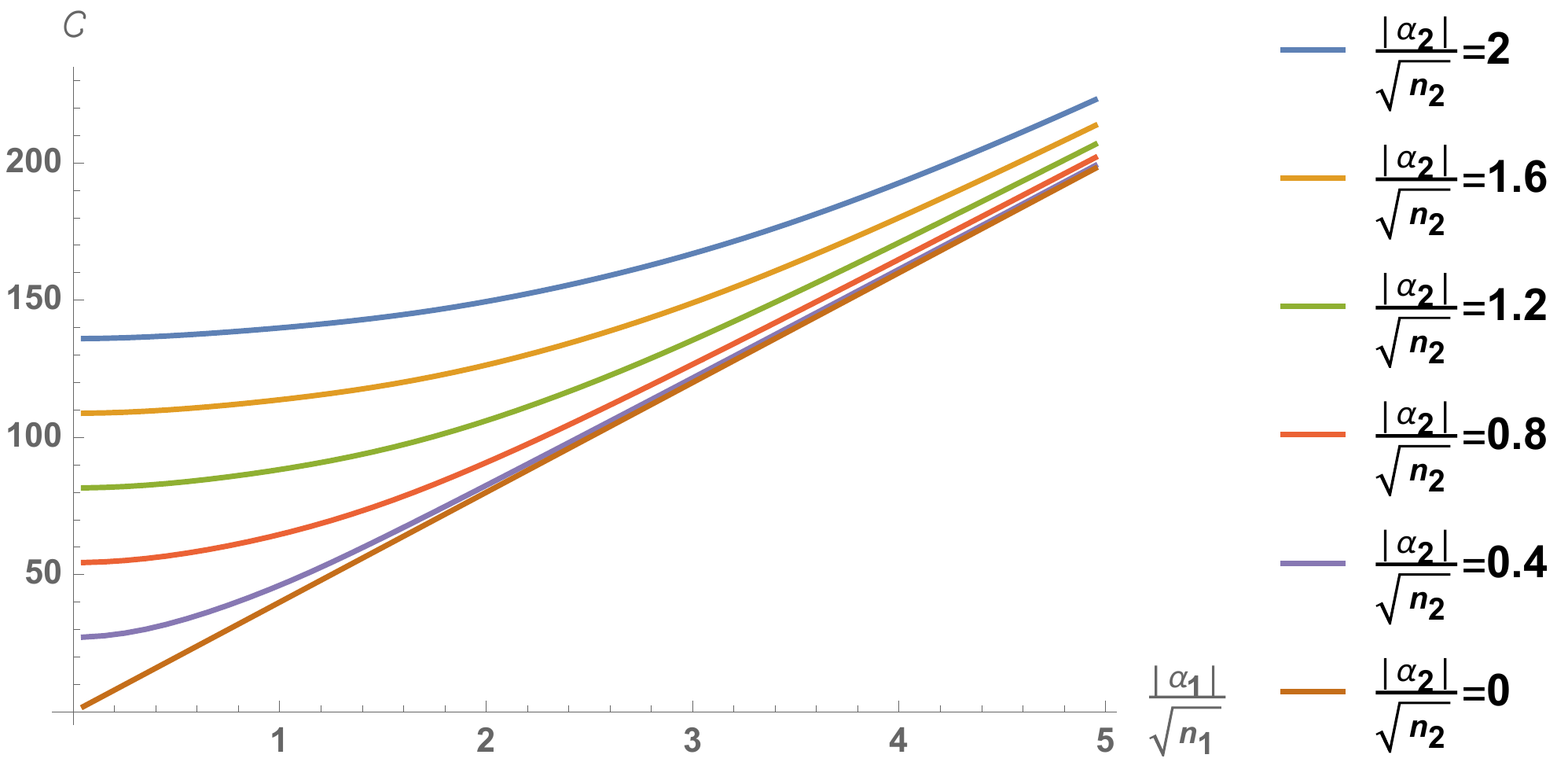}
	\caption{Two-mode shifts, $\mathcal{C}_{p=1}$}
	\label{fig:twomodeshift}
	\end{subfigure}
	\begin{subfigure}{0.48\linewidth}
 	\includegraphics[width=\linewidth]{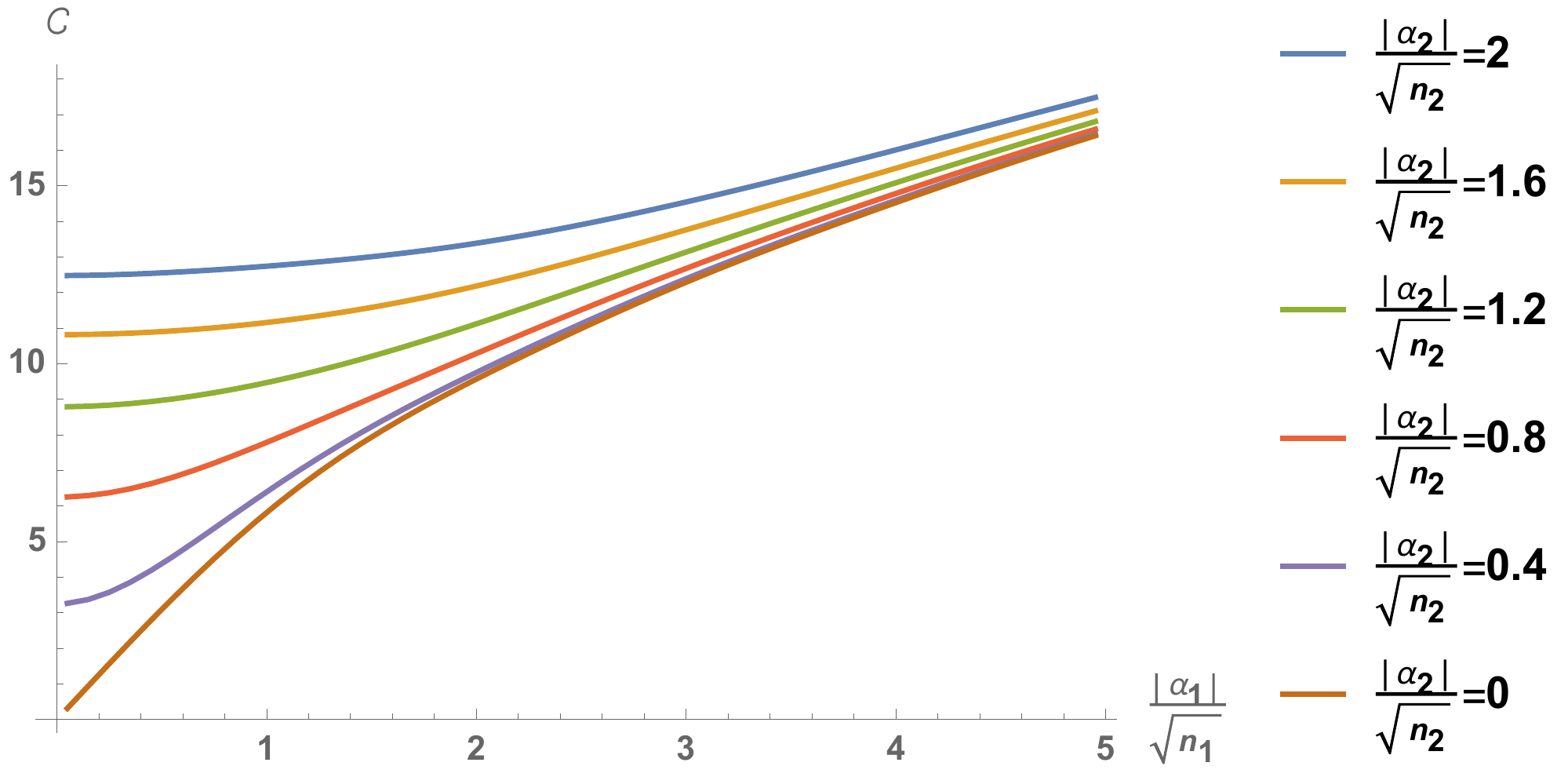}
	\caption{Two-mode shifts, $\mathcal{C}_{p=2}$}
	\label{fig:cohmultip2}
	\end{subfigure}
	\caption{Complexity  for two-mode shifts with $n_1=10, n_2=17$. The cutoff is chosen to be   $N=200$ for the two-mode shift covariance matrice, which is similar to one of the diagonal blocks in \eqref{eq:delta2}.}
\end{figure}

As has been considered in Section \ref{sec:bosongates}, one can also extend the result \eqref{eq:singlemodefermion} to shifts in more modes in terms of fermionic gates, \ie, the complexity between $\ket{\mathcal{G}_B^N}$ and $U_{n_1}(\alpha_1) \cdots U_{n_k}(\alpha_k)\ket{\mathcal{G}_B^N}$. Like for the single mode shift, the complexity would be independent of which bosonic ground state we are considering. Thus for simplicity as in the last section, one should focus on the case that $\ket{\mathcal{G}_B^N}$ is chosen to be the fermionic vacuum $\ket{0}$. The procedure is quite similar, first we have to obtain the transformation between the old and new basis in the following way,
\begin{align}
\hat{c}_l &= U^\dagger_{n_k}(\alpha_k)U^\dagger_{n_{k-1}}(\alpha_{k-1})\cdots U^\dagger_{n_1}(\alpha_1)c_l U_{n_1}(\alpha_1)U_{n_2}(\alpha_2)\cdots U_{n_k}(\alpha_k)\no\\
&= \sum_{j,m,\dots,i,q}R^{(k)}_{lj}(\alpha_k,n_k)R^{(k-1)}_{jm}(\alpha_{k-1},n_{k-1})\cdots  R^{(1)}_{iq}(\alpha_1,n_1)c_{q}\label{eq:basischangemulti}
\end{align}
where 
\begin{equation}
U^\dagger_{n_k}(\alpha_k)c_l U_{n_k}(\alpha_k)= \sum_j R^{(k)}_{lj}(\alpha_k,n_k)c_j,
\end{equation}
and the entries of $R^{(k)}(\alpha_k,n_k)$ can be read out from \eqref{eq:basisrel1} and \eqref{eq:basisrel2}. Since $U^\dagger_{n_k}(\alpha_k)$s are all commuting among themselves, $R^{(k)}(\alpha_k,n_k)$s forming a matrix representation are totally commuting as a result. In general, the arguments $\alpha_k$ are complex numbers $\alpha_k = |\alpha_k| e^{i\theta_k}$;  one overall phase can be reabsorbed with a redefinition of  $c_j$ and $\tilde{c}_j$ as in  \eqref{eq:phaseabsorbtion}. The numerical results we obtained for the two-mode shifts show that the complexity is in fact independent of 
both phases. We conjecture that this is true in general, although we cannot give a proof. 
\paragraph{Result with $p=1$ norm}Applying the method in the last section \ref{sec:onemodeferm} for two-mode shifts $U_{n_1}(\alpha_1)$ and $U_{n_2}(\alpha_2)$, one could obtain the complexity $\mathcal{C}_{p=1}$ plotted as in Fig. \ref{fig:twomodeshift} where $n_1$ and $n_2$ are taken to be $n_1=10$ and $n_2=17$.
 From the plot, we can see that for the $|\alpha_2| = 0$ curve, the complexity is a linear function of $|\alpha_1|$ with a proportionality constant being $4\sqrt{n_1}$. 
While for the other $|\alpha_2| \neq 0$ curves, we see that as $|\alpha_1|$ becomes dominant they asymptotes to be linear in $|\alpha_1|$ again with the same slope as for the curve $|\alpha_2| = 0$. Therefore, from the shape of the curves, we could deduce that the complexity for two-mode shift is 
 \begin{align}
 \mC_{p=1}(\ket{\mathcal{G}_B^N}, \, U_{n_1}(\alpha_1)U_{n_2}(\alpha_2)\ket{\mathcal{G}_B^N}) = 4\sqrt{n_1|\alpha_1|^2 + n_2|\alpha_2|^2},
 \end{align}
 which fits nicely with the results and reduces to \eqref{eq:singlemodefermion} when $|\alpha_2| = 0$. It is reasonable to suppose that the result will generalize to multi-mode shifts as follows: 
\begin{align}\label{eq:multimodefermion}
\mC_{p=1}(\ket{\mathcal{G}_B^N}, \, U_{n_1}(\alpha_1)U_{n_2}(\alpha_2)\cdots U_{n_k}(\alpha_k)\ket{\mathcal{G}_B^N}) = 4\sqrt{\sum_{i=1}^{k}n_i|\alpha_i|^2} \,.
\end{align}

\paragraph{Result with $p=2$ norm} We plot the complexity $\mC_{p=2}$ with the same setup in Fig. \ref{fig:cohmultip2}.  As in Fig. \ref{fig:twomodeshift}, the complexity increases as either $\alpha_1$ or $\alpha_2$ increases, and all the curves converge to a certain curve as $\frac{|\alpha_1|}{\sqrt{n_1}}$ becomes much larger than $\frac{|\alpha_2|}{\sqrt{n_2}}$, which is expected since then the effect of the first shifting mode becomes dominant.


Finally, the upper bound for the complexity with \textit{Schatten} $p=2$ norm is given by
\begin{equation}
\mC_{p=2}(\ket{\mathcal{G}_B^N} ,  U_{n_1}(\alpha_1)U_{n_2}(\alpha_2)\ket{\mathcal{G}_B^N}) \le 4\sqrt{n_1|\alpha_1|^2 + n_2|\alpha_2|^2} \,,
\end{equation}
and similarly for the general case. 

\section{A class of fermionic and bosonic Gaussian states}\label{sec:gaussequiv}
In the previous sections, we have shown that the bosonic coherent states of the type in \eqref{eq:coherentexample} are fermionic gaussian states and we have also seen how to obtain the corresponding complexity by using Fubini-Study method and Nielsen method in terms of bosonic gates and fermionic gates respectively. 
However the bosonic coherent states are only a subset of all the fermionic gaussian state, whereas a state that is fermionic gaussian is not generically also bosonic gaussian. 

In this section we show that there is a large class of states that are simultaneously bosonic and fermionic gaussian. This allows us to study the effect that different choices of gates have on the complexity of a given state, since in the fermionic description we allow gates that are fermion bilinears, whereas in the bosonic description we use gates that are bosonic bilinears.  Notice that neither set of gates is a subset of the other, so we cannot establish a priori a  bound between the two complexities.

\subsection{A bosonisation identity}

Starting with the fundamental bosonisation formula  \eqref{eq:fieldidentity}, it is possible to obtain the following relation \cite{reviewbosonisation}:
\begin{align}\label{eq:fieldoperator}
\normord{i\psi^\dagger(x) \der_x \psi(x)} = {2\pi^2 \over L^2} (\hat{N} + \hat{N}^2) + {\pi \over L} \der_x\phi(x) + {1\over 2} \normord{(\der_x\phi(x))^2} + {2 \pi \over L} \hat{N} \der_x \phi(x) + {i \over 2} \der_x^2\phi(x) \,.
\end{align}
Taking the integral over space of this relation, we obtain on the two sides the fermionic and bosonic hamiltonian respectively, which is the basic property of the bosonisation formalism. 
Formula \eqref{eq:fieldoperator} shows that the equality holds at the level of the hamiltonian density, up to total derivative terms which, crucially for our purposes, are 
linear in the boson field, hence fermion bilinears, if we work on a subspace of fixed particle number. For definiteness we can consider the case $\hat{N}=0$, \ie,  the module with fermion number zero. Therefore if we consider the inhomogeneous hamiltonian $H_f = \int dx f(x)  \normord{i\psi^\dagger(x) \der_x \psi(x)}$, dependent on an arbitrary function $f(x)$, the corresponding ground state will be fermionic gaussian, and also bosonic gaussian. 

It is unlikely that further identities of this type exist, therefore we conjecture that the states obtained this way are the only ones that are gaussian in both descriptions.  If we write down the same relation in Fourier components, taking 
\begin{equation}
f(x) = \sum_k f_k e^{i {2\pi k x\over L}}
\end{equation}
and integrating each term of  \eqref{eq:fieldoperator} multiplied by $f(x)$ over the whole space $[-L/2,L/2]$ gives 
\begin{align}
&\int_{-L/2}^{L/2} {dx\over 2\pi} f(x) \der_x \phi(x) = - i\sum_{n>0} \sqrt{n}(f_n b_n^\dagger - f_{-n} b_n),\\
&\int_{-L/2}^{L/2} {dx\over 2\pi} f(x) \normord{(\der_x \phi(x))^2 } = -{2\pi \over L} \sum_{m,n>0} \sqrt{mn} (f_{m+n} b_m^\dagger b_n^\dagger +f_{-m-n} b_m b_n - 2f_{n-m}b_n^\dagger b_m),\\
&\int_{-L/2}^{L/2} {dx\over 2\pi} f(x) (i\der_x^2\phi(x)) = {2\pi i \over L}\sum_{n>0} n^{3/2}(f_n b_n^\dagger +f_{-n} b_n),\\
&\int_{-L/2}^{L/2} {dx\over 2\pi} f(x) \normord{i\psi^\dagger(x)\der_x \psi(x)} = {2\pi  \over L}\sum_{l,k} k f_{l-k} \normord{c^\dagger_l c_k} = {2\pi  \over L} \sum_{l,k}k\normord{c^\dagger_{k+l}c_k} f_l,
\end{align}
putting all terms together we have the following relation 
\begin{multline}\label{eq:modesidentity}
\sum_l l \normord{c_l^\dagger c_l}f_0 +\sum_{n\neq0,l} \left(l+\frac{n-2\hat{N} -1}{2} \right) c_{l+n}^\dagger c_l f_n \\= {1 \over 2}(\hat{N} + \hat{N}^2) f_0  - {1\over 2}\sum_{m,n> 0} \sqrt{mn} (f_{m+n} b_m^\dagger b_n^\dagger +f_{-m-n} b_m b_n - 2f_{n-m}b_n^\dagger b_m) \,.
\end{multline}
Once again we see that identifying the coefficient of $f_0$, we obtain the usual identity for fermionic and bosonic Hamiltonians
\begin{equation}
\sum_{k} k\normord{c^\dagger_{k}c_k}  = \sum_{m>0} m b_m^\dagger b_m + \frac{1}{2} (\hat{N} + \hat{N}^2).
\end{equation}

 \subsection{An example with one mode}
We consider now one particular example of states in this family, corresponding to a function that has only Fourier modes $n= \pm 2$. 
The operator identity for  $f_2$ which can be read from eq. \eqref{eq:modesidentity} is 
\begin{equation}
\sum_l \left( l+ {1-2 \hat{N} \over 2} \right) c_{l+2}^\dagger c_l = - {1\over 2} b_1^\dagger b_1^\dagger + \sum_{m>0} \sqrt{(m+2)m} \, b_{m+2}^\dagger b_m ,
\end{equation}
and for $f_{-2}$ we have the hermitian conjugate of the above. 
The two identities will help build the following unitary operator $W_f(\beta) = W_b(\beta)$, given in terms of fermionic modes and bosonic modes, respectively, as
\begin{align}
W_f(\beta)  &= \textrm{exp}\left( \sum_l \left( l+ {1 - 2 \hat{N} \over 2} \right)(\beta c_{l+2}^\dagger c_l - \beta^* c_{l}^\dagger c_{l+2}) \right) \,,\\
W_b(\beta) &= \textrm{exp}\left( -{\beta \over 2}b_1^\dagger b_1^\dagger + {\beta^* \over 2}b_1 b_1 + \sum_{m>0} \sqrt{m(m+2)} \, (\beta b_{m+2}^\dagger b_m - \beta^* b_{m}^\dagger b_{m+2}) \right). 
\end{align}

We will assume $\beta\in \mathbb{R}$ and focus on the sector $\hat{N}=0$ in the following part with the reference state being the fermionic vacuum $\ket{0}$. The adjoint action of these operators on the modes cannot be obtained in an analytically closed form as in \eqref{eq:basisrel1}, \eqref{eq:basisrel2}. We must resort to a fully numerical computation. 
We put a cutoff on the number of modes, find the matrices representing  ${1\over \beta}\ln W_f$ and ${1\over \beta}\ln W_b$, respectively $R_f$ and $R_b$, and compute the  exponential of those two matrices, which give the change of basis $\tilde \xi_i^f = (e^{\beta R_f})_{ij} \xi_j^f$ and $\tilde \xi_i^b = (e^{\beta R_b})_{ij} \xi_j^b$. The relative covariance matrix is then computed as in ~\eqref{relative-covariance},\eqref{relative-covariance-bos}, and the eigenvalues are found numerically. 
There is one case when we can find an exact result: the bosonic covariance $\Delta_{b}$ has  the form $\Delta_{b}(\beta)=e^{\beta R_b}e^{\beta R_b^T}$, and
\begin{align}\label{eq:bosongaussian}
 R_b = \begin{pmatrix}
A_b & S_b\\
S_b & A_b
\end{pmatrix}\sim \begin{pmatrix}
A_b + S_b&0\\
0 & A_b -S_b
\end{pmatrix}
\end{align}
where $A_b$ is antisymmetric and $S_b$ is symmetric, which is consistent with the fact that $\Delta_b  \in Sp(2N)$. Equation \eqref{eq:bosongaussian}  means that $\ln\Delta_b$ can always be block diagonalized as
\begin{align}\label{eq:bosongaublock}
\ln\Delta_b (\beta)= \begin{pmatrix}
\ln\left( e^{\beta(A_b + S_b)}e^{\beta(S_b -A_b) } \right) &0\\
0& \ln\left( e^{\beta(A_b - S_b)}e^{\beta(-A_b -S_b) } \right) 
\end{pmatrix}.
\end{align}
We notice that if one of the two blocks is positive definite, the other would be negative definite, due to the fact that the eigenvalues of $\Delta_b$ always come in pairs $e^{\pm r}$ (this is because $\Delta_b$ is symplectic and orthogonal).  In this case one can show that the antisymmetric part does not contribute to the complexity; then it is easy to see that, setting $A_b=0$ in \eqref{eq:bosongaublock}, 
the complexity $\mathcal{C}_{p=1}$ can be computed exactly and is linear in the trace of the symmetric part $S_b$. However we do not have a criterion to determine when the blocks will be postive definite. 

\begin{figure}[h!]
  \centering
   \begin{subfigure}[b]{0.45\linewidth}
    \includegraphics[width=\linewidth]{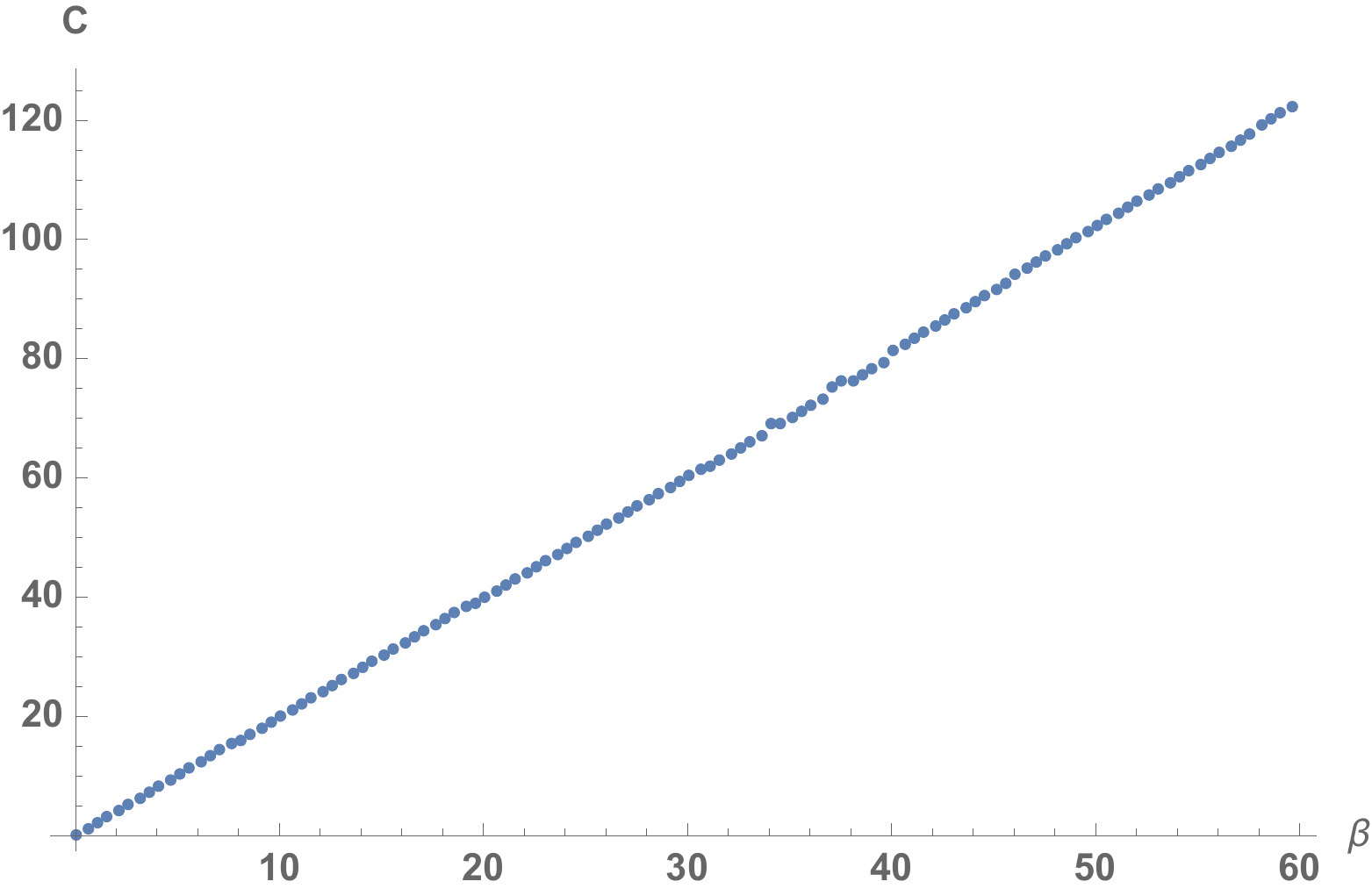}
    \caption{Boson, $\mC_{p=1}$, $N=40$}
    \label{fig:bosonN40}
  \end{subfigure}
   \begin{subfigure}[b]{0.45\linewidth}
    \includegraphics[width=\linewidth]{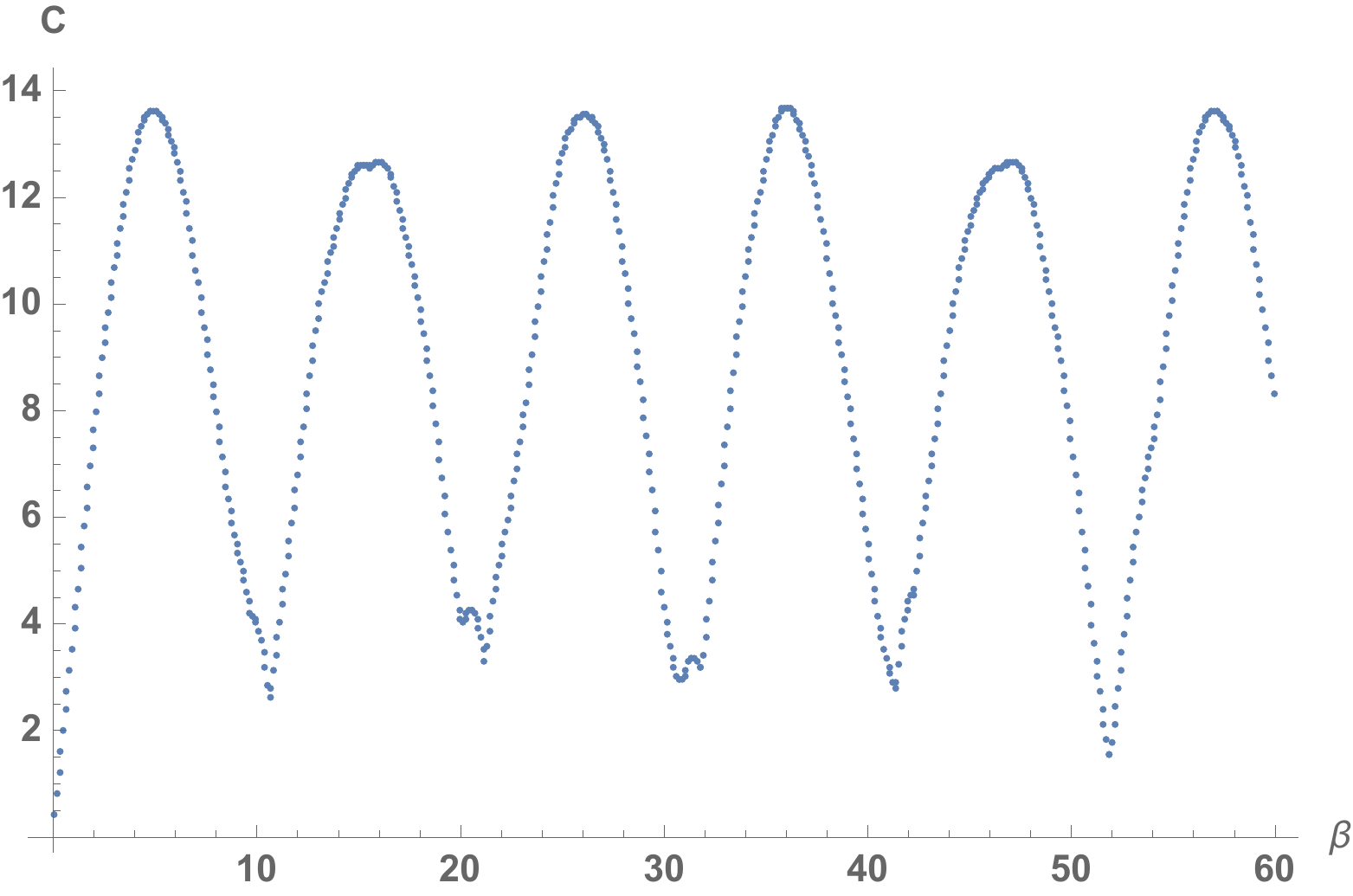}
    \caption{Fermion, $\mC_{p=1}$, $N=40$}
    \label{fig:fermionN40}
  \end{subfigure}
    \begin{subfigure}[b]{0.45\linewidth}
    \includegraphics[width=\linewidth]{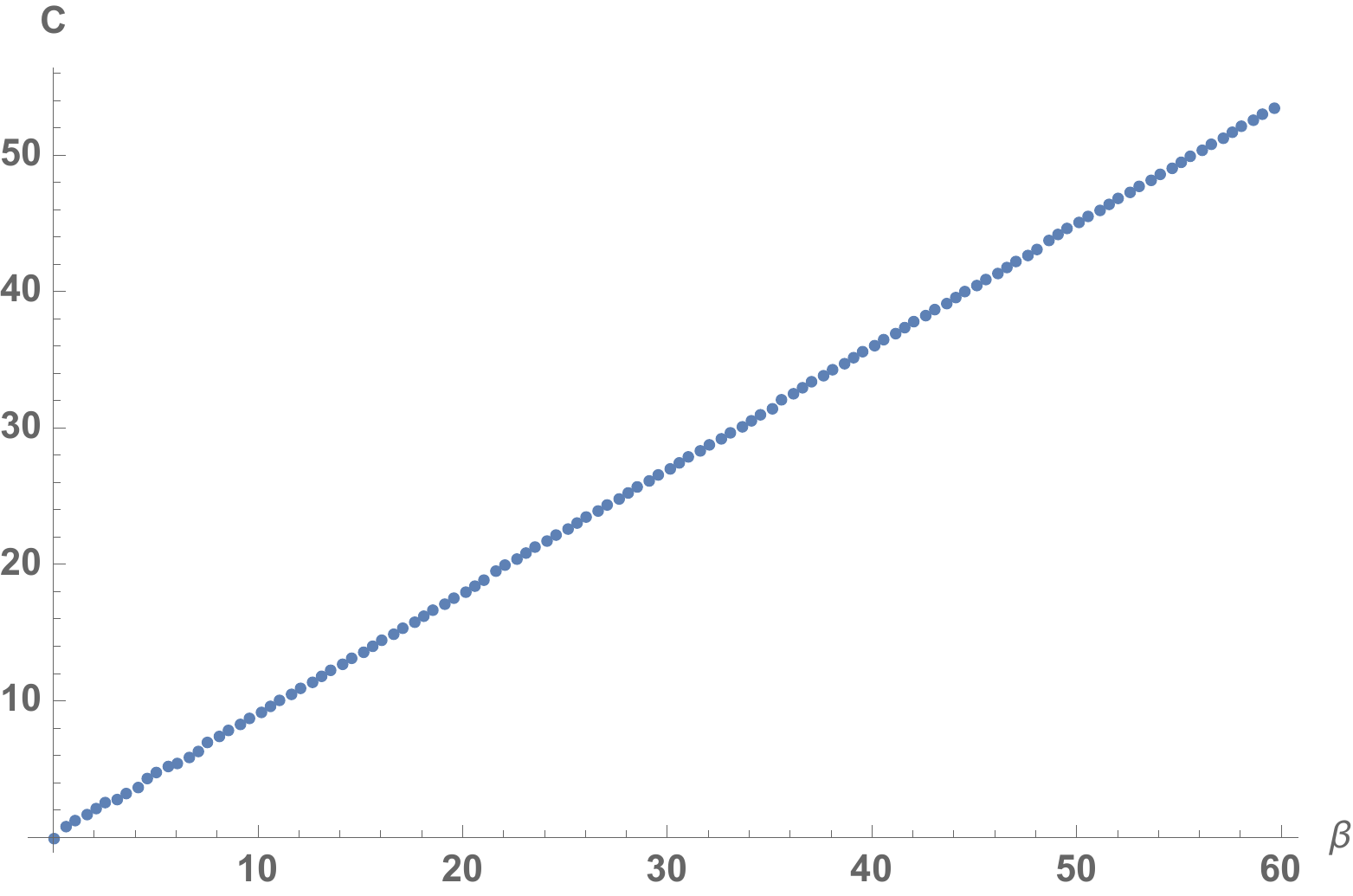}
    \caption{Boson, $\mC_{p=2}$, $N=40$}
    \label{fig:bosonN40p2}
  \end{subfigure}
   \begin{subfigure}[b]{0.45\linewidth}
    \includegraphics[width=\linewidth]{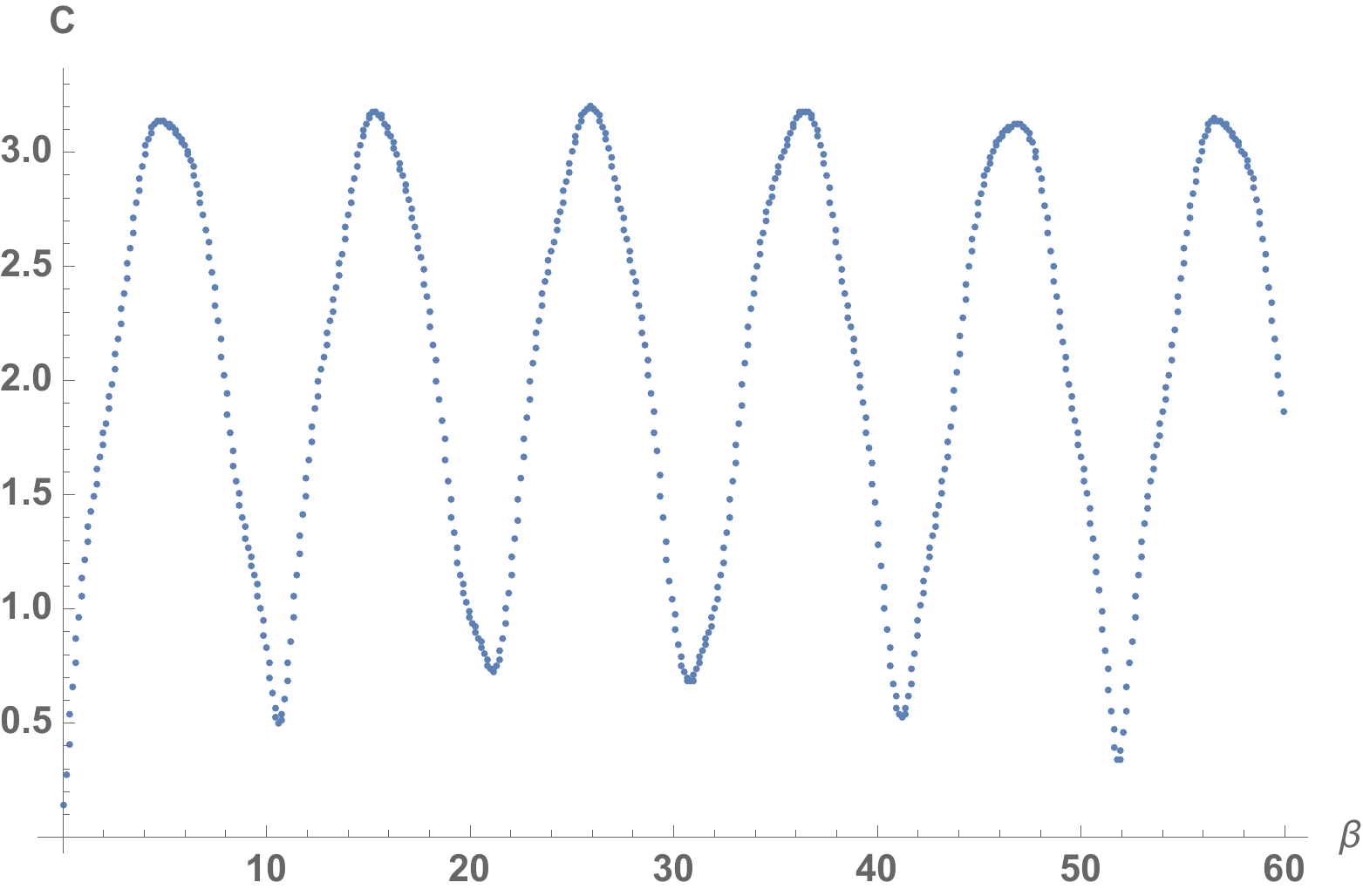}
    \caption{Fermion, $\mC_{p=2}$, $N=40$}
    \label{fig:fermionN40p2}
  \end{subfigure}
  \caption{The complexity for bosonic and fermionic gaussian states with $p=1$ and $p=2$ norms. (a) and (c) represent the bosonic case while (b) and (d) represent the fermionic case. The cutoff is chosen to be $N=40$.  }
  \label{Level-two-bigauss}
\end{figure}

\begin{figure}[h!]
\centering
  \begin{subfigure}[b]{0.45\textwidth}
    \includegraphics[width=\linewidth]{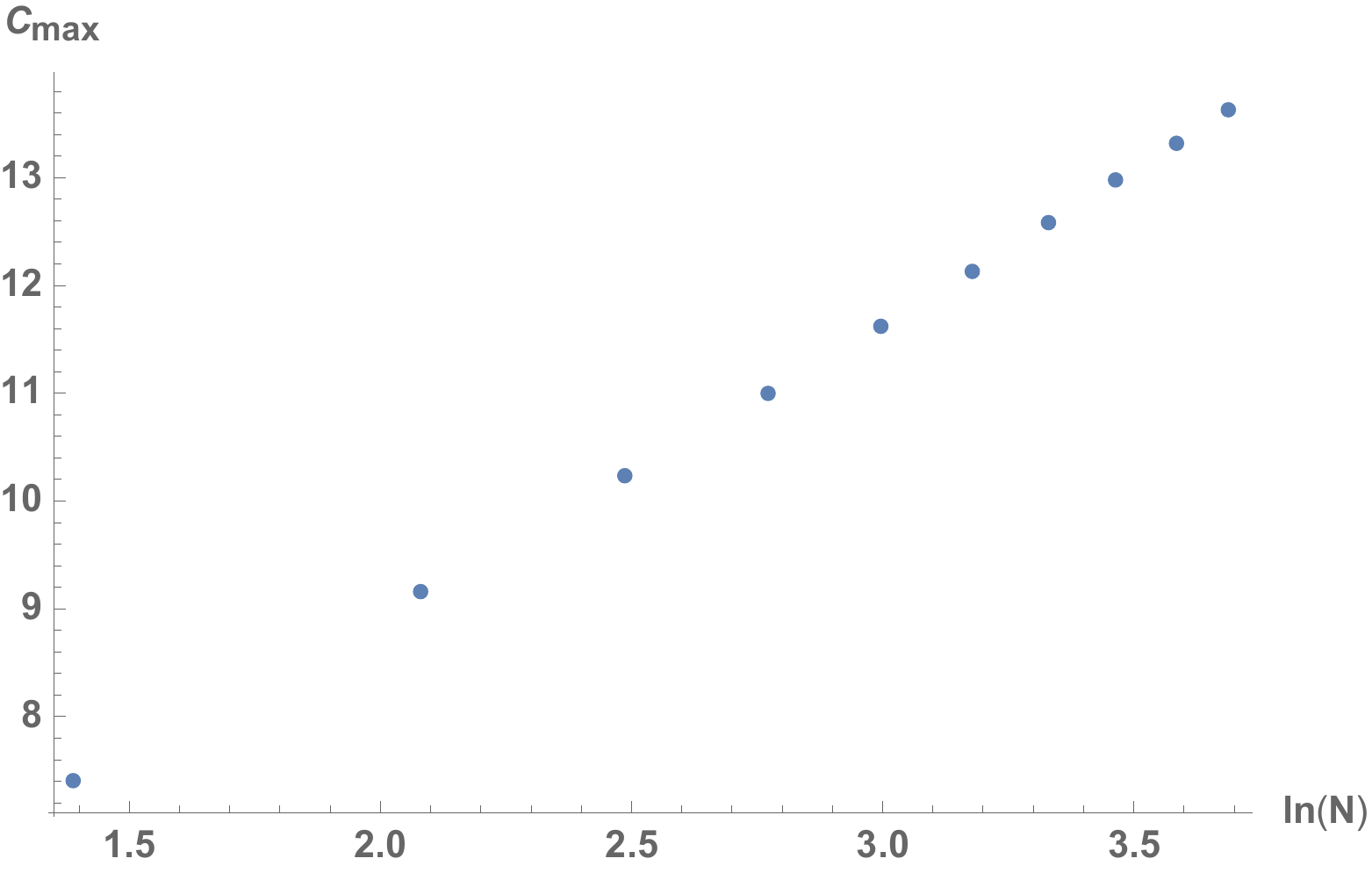}
    \caption{$\mC_{p=1}$ v.s. $\ln(N)$}
    \label{fig:fmaxp1}
  \end{subfigure}
   \begin{subfigure}[b]{0.45\linewidth}
    \includegraphics[width=\linewidth]{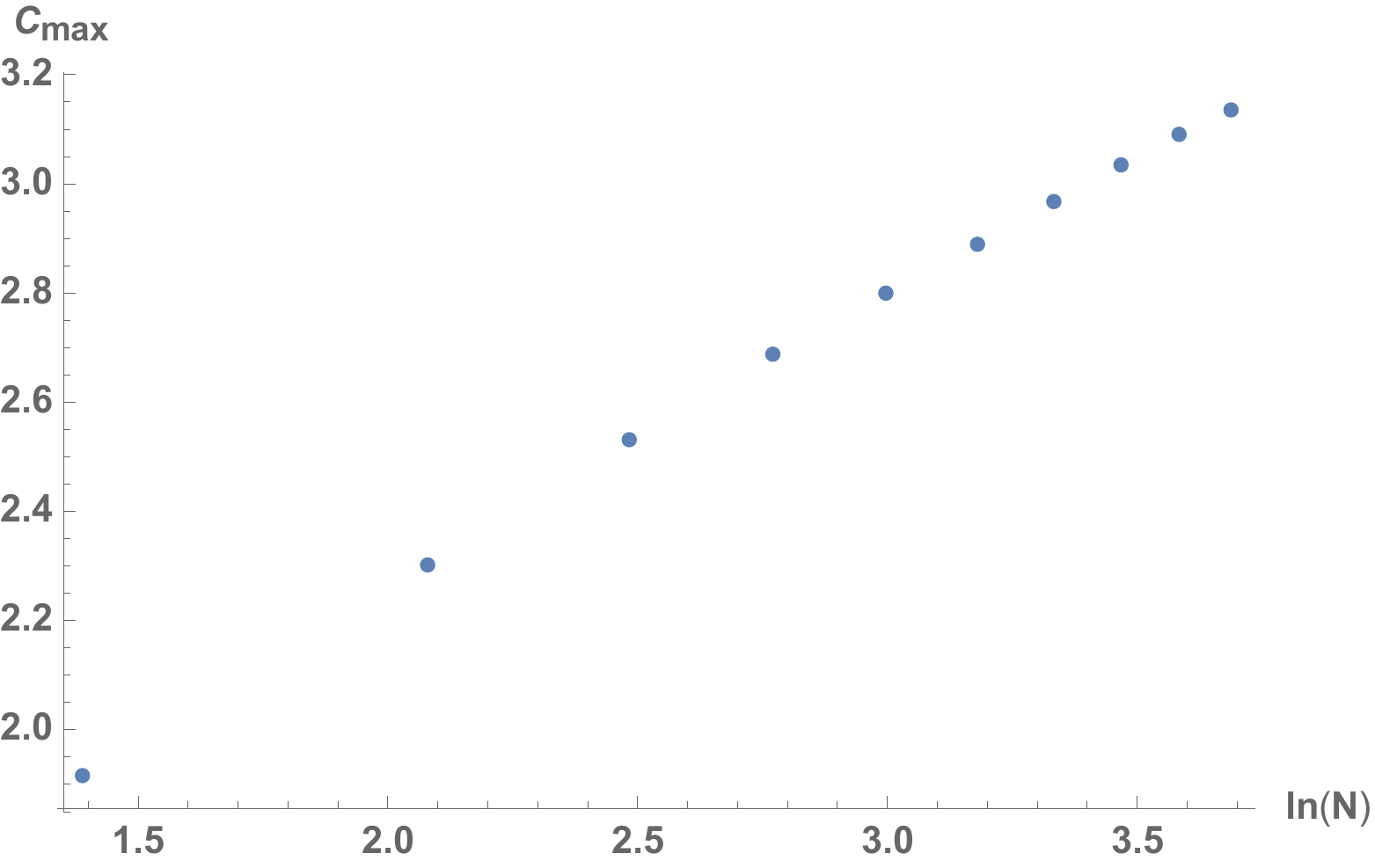}
    \caption{$\mC_{p=2}$ v.s. $\ln(N)$}
    \label{fig:fmaxp2}
  \end{subfigure}
  \caption{The maximum complexity for the fermionic case grows w.r.t. the logarithm of the cutoff, linearly for the $p=1$ norm and quasi-linearly for the $p=2$ norm.}
    \label{fig:Fermion-level-two-cutoff}
\end{figure}

\paragraph{Result with $p=1$ norm}The results are shown in Fig. \ref{Level-two-bigauss}. The behavior appears to be very different for bosons and fermions.  
 The bosonic complexity (see Fig. \ref{fig:bosonN40}) grows linearly without bound; this linear growth can be explained, as discussed above, by the positive-definiteness of the blocks of the covariance matrix. 
 On comparison, the fermionic complexity is oscillating (see Fig. \ref{fig:fermionN40}) in a way that suggests a quasi-periodic function. The maximum complexity of the first peak seems to grow linearly w.r.t. the logarithm of the cutoff $\ln(N)$ as in Fig. \ref{fig:fmaxp1}, this is consistent with the fact that the orthogonal group $SO(2N)$ becomes non-compact as $N\to \infty$.  For bosons, on the other hand, this particular class of states has a complexity that is independent of the cutoff, even though the transformation at finite $\beta$ mixes all the modes among themselves and not just a finite number of them. 

\paragraph{Result with $p=2$ norm} The complexities for gaussian states complexity with $p=2$ norm are plotted in Fig. \ref{fig:bosonN40p2} for the bosonic case and in Fig. \ref{fig:fermionN40p2} for fermionic one. The trends appear to be similar as the $p=1$ case, except that the values of the complexity are smaller. 

\subsection{An example with two modes}
\begin{figure}[h!]
  \centering
   \begin{subfigure}[b]{0.45\linewidth}
    \includegraphics[width=\linewidth]{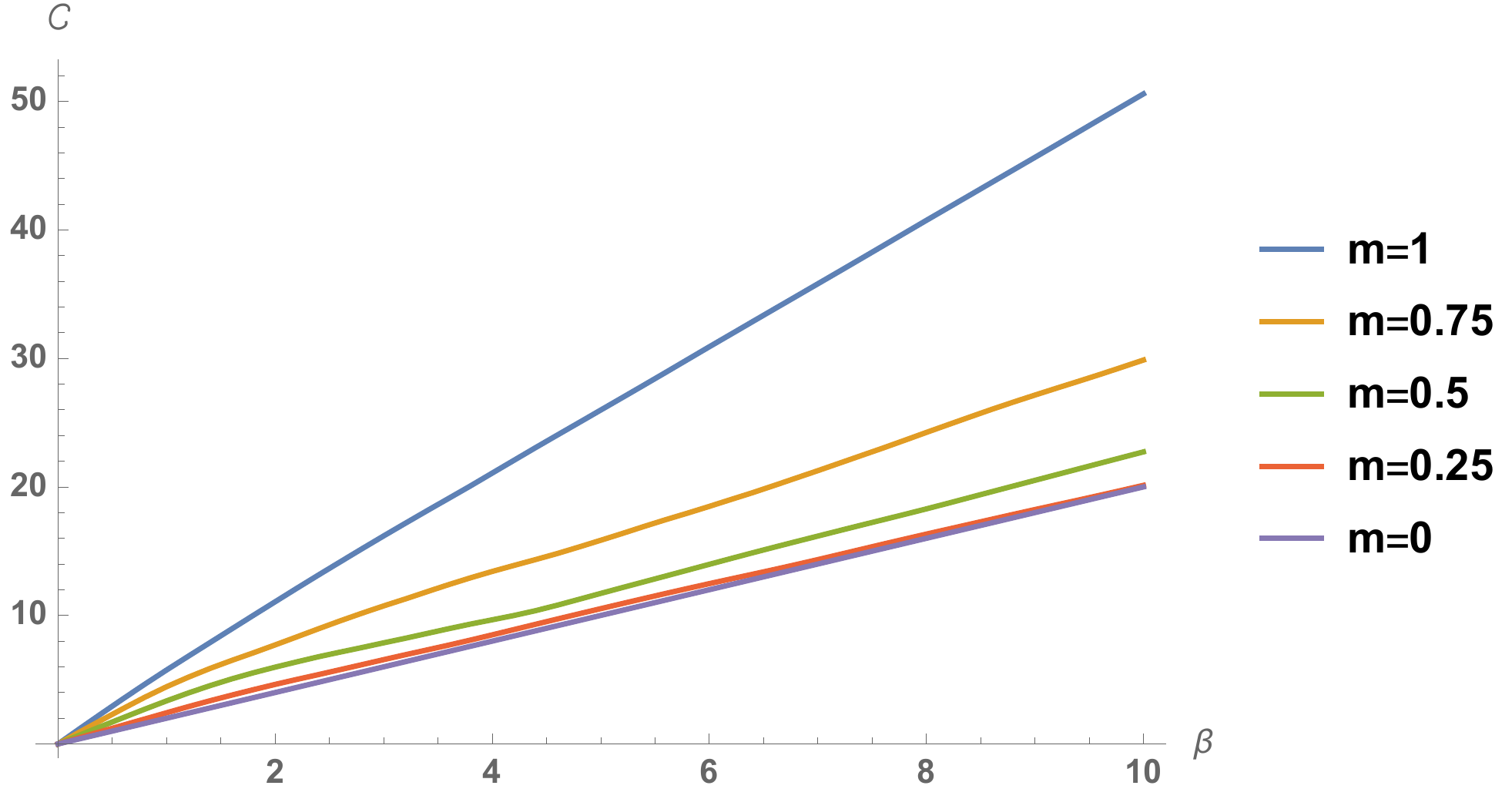}
    \caption{Boson, $\mC_{p=1}$, $N=24$}
    \label{fig:btwomodeN24p1}
  \end{subfigure}
   \begin{subfigure}[b]{0.45\linewidth}
    \includegraphics[width=\linewidth]{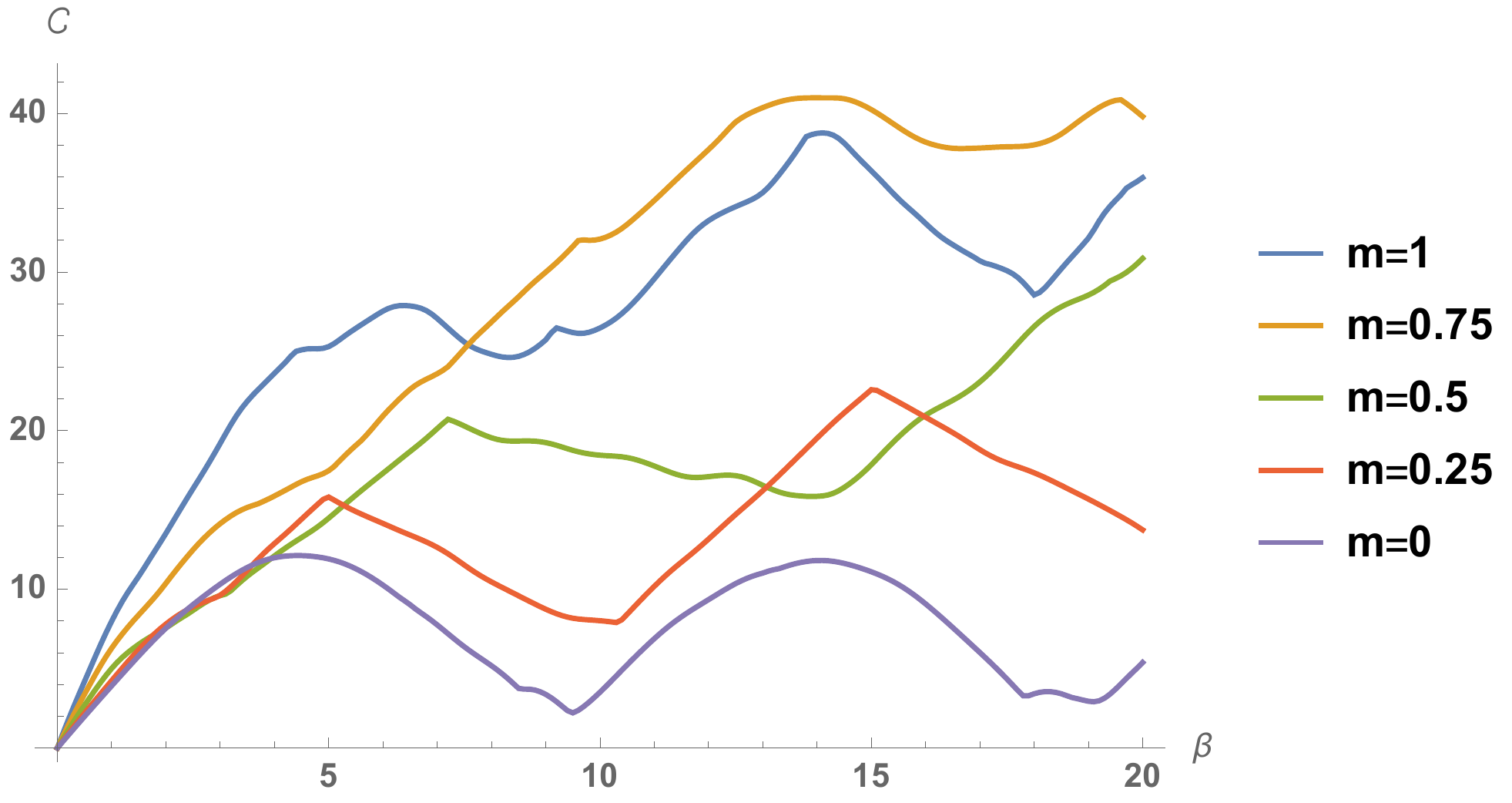}
    \caption{Fermion, $\mC_{p=1}$, $N=24$ }
    \label{fig:ftwomodN24p1}
  \end{subfigure}
    \begin{subfigure}[b]{0.45\linewidth}
    \includegraphics[width=\linewidth]{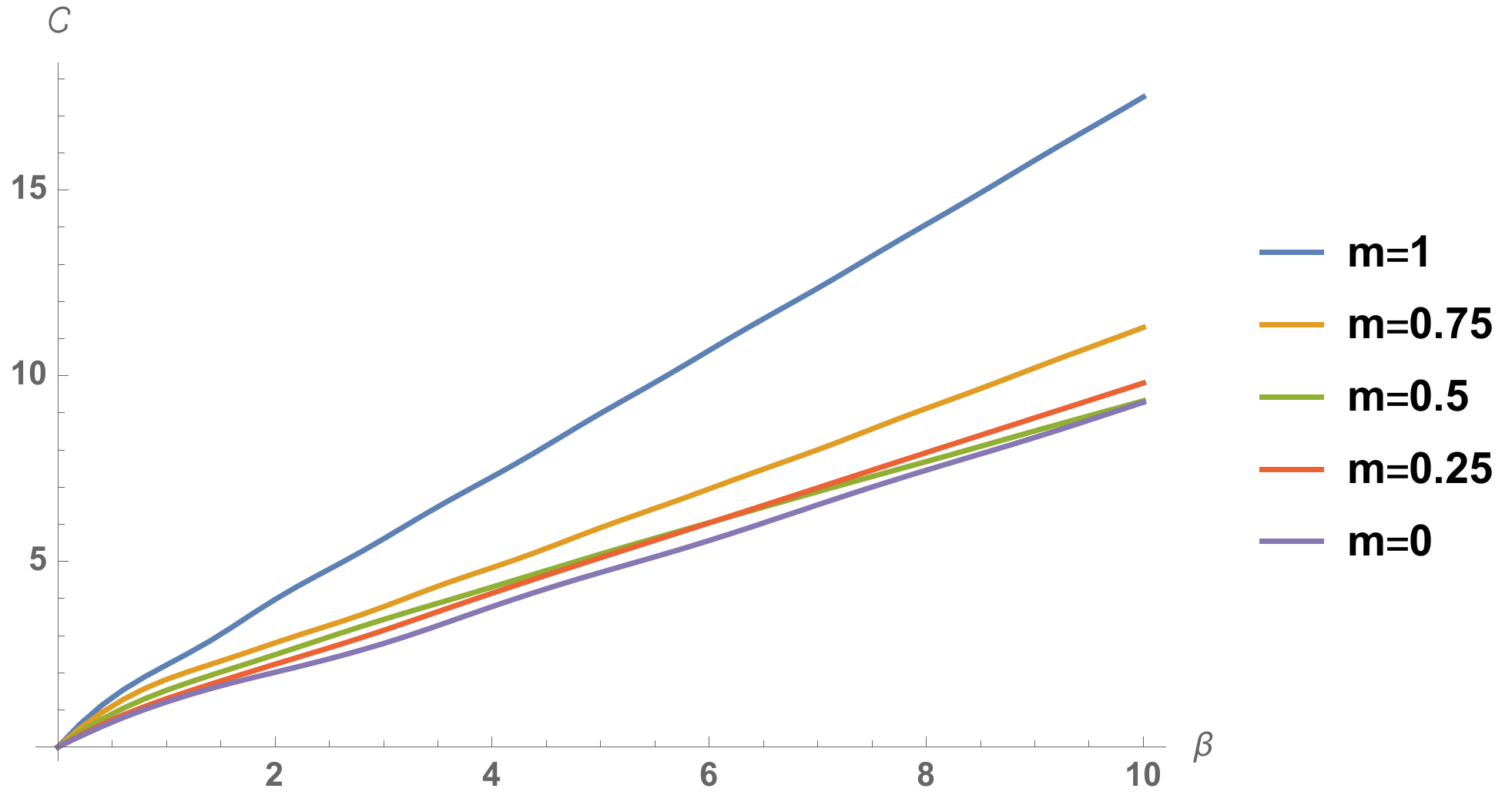}
    \caption{Boson, $\mC_{p=2}$, $N=24$ }
    \label{fig:btwomodeN24p2}
  \end{subfigure}
   \begin{subfigure}[b]{0.45\linewidth}
    \includegraphics[width=\linewidth]{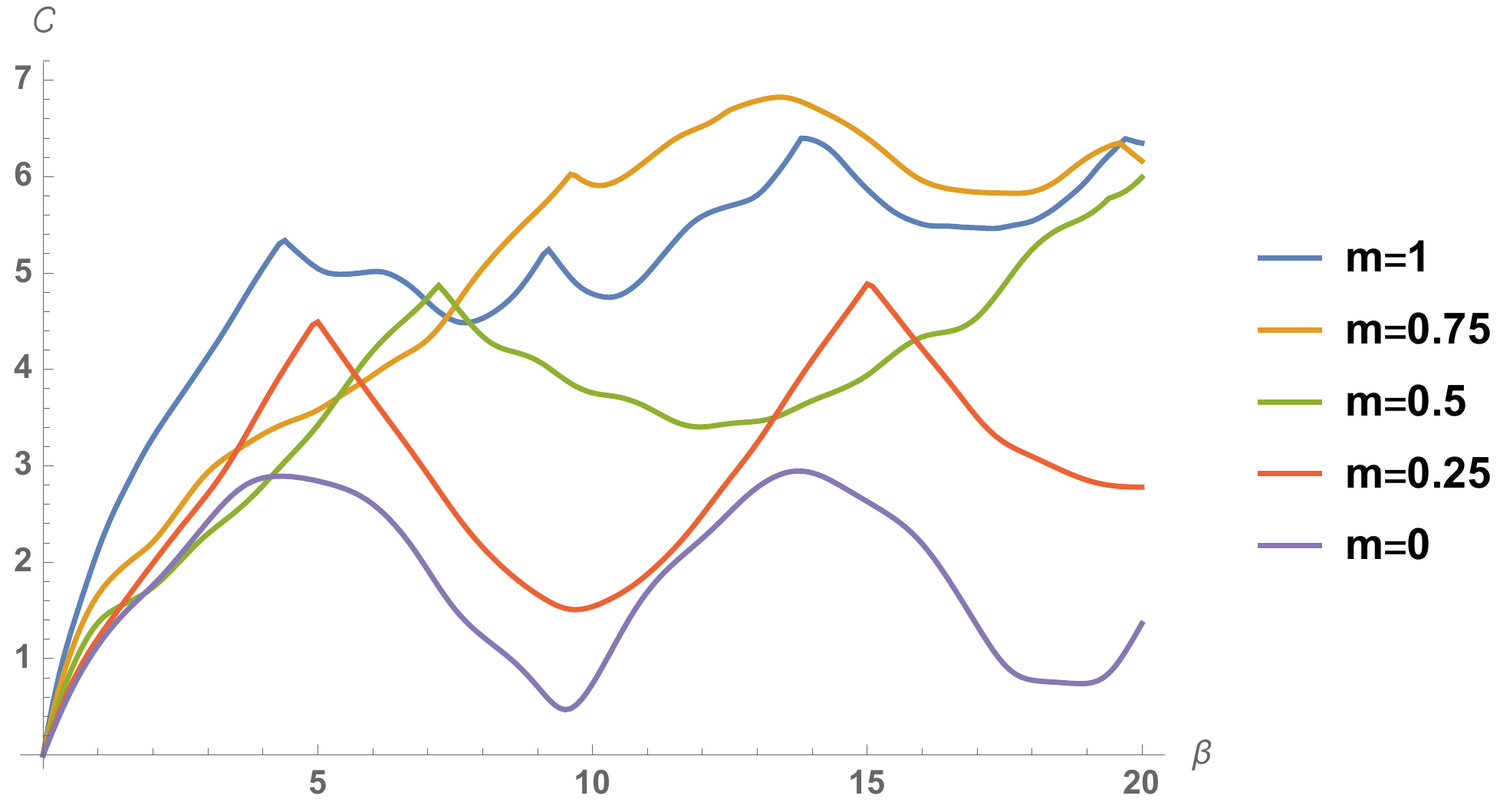}
    \caption{Fermion, $\mC_{p=2}$, $N=24$}
    \label{fig:ftwomodN24p2}
  \end{subfigure}
  \caption{The complexity for bosonic and fermionic gaussian states with $p=1$ and $p=2$ norms. (a) and (c) represent the bosonic case while (b) and (d) represent the fermionic case. The cutoff is chosen to be $N=24$ and the ratio varies from $0$ to $1$ with an interval of $0.25$.  }
  \label{fig:bigaussiantwo}
\end{figure}

We consider next  a function $f(x)$ with two modes;  the simplest case would be to use the operator identities for $f_{\pm 2}$ and $f_{\pm 3}$. In this case, the unitary operator for fermions 
\begin{align}
W_f(\beta,\gamma)  = \textrm{exp}\Bigg[& \sum_l \left( l+ {1 - 2 \hat{N} \over 2} \right)(\beta c_{l+2}^\dagger c_l - \beta^* c_{l}^\dagger c_{l+2}) \no\\
+ &\sum_l \left( l+ 1 - \hat{N} \right)(\gamma c_{l+3}^\dagger c_l - \gamma^* c_{l}^\dagger c_{l+3})\Bigg] \,,
\end{align}
 and for bosons
 \begin{align}
W_b(\beta,\gamma) = \textrm{exp}\Bigg[& -{\beta \over 2}b_1^\dagger b_1^\dagger + {\beta^* \over 2}b_1 b_1 + \sum_{m>0} \sqrt{m(m+2)} \, (\beta b_{m+2}^\dagger b_m - \beta^* b_{m}^\dagger b_{m+2})\no\\
& -\sqrt{2}\gamma b_1^\dagger b_2^\dagger + \sqrt{2} \gamma^* b_1b_2 + \sum_{m>0} \sqrt{m(m+3)} \, (\gamma b_{m+3}^\dagger b_m - \gamma^* b_{m}^\dagger b_{m+3})\Bigg] \,,
\end{align}
depend on two parameters, $\beta$ and $\gamma$, coupling to the $n=2$ and $n=3$ mode respectively. 
For simplicity,  we consider the case where $\beta$ and $\gamma$ are real and proportional to each other as $\gamma= m\beta$. In Fig. \ref{fig:bigaussiantwo}, we plotted the complexity  for both the bosonic state (Fig. \ref{fig:btwomodeN24p1} and \ref{fig:btwomodeN24p2}) and the fermionic state (Fig. \ref{fig:ftwomodN24p1} and \ref{fig:ftwomodN24p2}) as a function of $\beta$ and for various values of $m$. We notice that in the bosonic case, the mixing of the modes does not have a dramatic effect (although the curves are no longer exactly linear), while for fermions the quasi-periodic behavior is destroyed. 
Again we observe that the qualitative features are the same using the  $p=1$ and the $p=2$ norm.

%

\section{Conclusions}\label{Conclusions}
In this paper, we made a first step to consider the relations between fermionic and bosonic field theory complexity. We investigate such relations in the $2D$ free boson/free fermion model, where a highly nonlinear exact relation between the scalar field and the fermion field is present, encoded in eq. \eqref{eq:fieldidentity} that expresses  the fermion field as the vertex operator of the scalar field. In this setup, the fermonic Fock space can be decomposed into modules according to their fermionic number (see eq. \eqref{eq:fockspace}), and each module can be identified with the bosonic Hilbert space. Moreover, the space of quadratic operators in one description is mapped to a 
space of non-quadratic ones in the other, and this allows us to study how the choice of the set of allowed gates influences the complexity of an operator. 

In the current framework of calculable field theory complexity, not all the states can be treated on equal footing. We identified two classes of states that can be considered from both sides. The first type is bosonic-coherent and fermionic-gaussian, a direct consequence of eq. \eqref{eq:bosonfield} where a single bosonic operator is a sum of bilinears of fermionic ones. We applied the \textit{Fubini-Study} metric method on the bosonic side, with the analytical results given in \eqref{eq:bosonresultOM} for a single mode shift and in \eqref{eq:twoshift} a generalization to multi-mode shift. The general multi-mode result shows that different modes are orthogonal in the  FS metric which is a flat Euclidean metric. On the fermionic side, we applied instead the \textit{Nielsen} method developed in \cite{Hackl:2018ptj}, with numerical result plotted in Fig. \ref{fig:largeN}, \ref{fig:twomodeshift} for \textit{Schatten} $p=1$ norm, and Fig. \ref{fig:cohp2}, \ref{fig:cohmultip2} for $p=2$ norm. Based on those numerics, an analytical result for multimode shift with $p=1$ norm is obtained in eq. \eqref{eq:multimodefermion}, which differ from the bosonic result  \eqref{eq:twoshift} by an extra dependence on the mode number. This means that if we assign a mode-number dependent cost in the bosonic case, the two results would be equivalent up to a total factor. This result is a bit surprising, since it is for the $p=2$ norm that we can interpret the complexity in terms of Riemannian geometry, so we should expect that the FS metric would compare more directly to that case. 
The $p=1$ norm gives the space the structure of a \textit{Finsler} manifold. It would be interesting to understand better the structure of this geometry; this could also shed some light on the 
relation between bosonic states and fermionic states holographically, since bosonic coherent states in the bulk constitute a sector of states that can be explcitly described both in the bulk and in the boundary (see the recent work  \cite{Bernamonti:2019zyy}). 

The second class of states is identified through the relation \eqref{eq:fieldoperator} which is a local (in space) relation between bosonic bilinears and fermionic ones, thus relating the gaussian states on both sides. These states can be understood as the ground states for an inhomogeneous Hamiltonian, obtained integrating \eqref{eq:fieldoperator} with an arbitrary function $f(x)$, that has the interpretation of a space-dependent Fermi  velocity. We obtain a one-function family of bigaussian states, out of which we studied some of the simplest cases using \textit{Nielsen} method. 
For the case of a function with the only Fourier component $f_{\pm2} = \alpha$, the numerical results for bosons and fermions are plotted in Fig. \ref{Level-two-bigauss}, which show qualitatively the same behavior using the  $p=1$ and $p=2$ norm. The most notable features are: the bosonic complexity grows linearly with $\alpha$, and is independent of the UV cutoff, while the fermionic complexity is roughly periodic in $\alpha$, reaching peaks with value scaling like $\ln N$ in the cutoff. This shows that the states are special in some sense, since a generic point in the group manifold $SO(2N)$ would be at a distance $\sim \sqrt{N}$ from the identity (at least in the $p=2$ norm).
We also considered the next simplest two-mode case, with components $f_2$ and $f_3$. The results, plotted in Fig. \ref{fig:bigaussiantwo}, show that the effect of the mixing of the modes is small on the bosonic complexity, but more dramatic one on the fermionic one. It would be interesting to understand what is the behavior for a generic function $f$. 

Incidentally, it would be also interesting to know whether these bi-Gaussian states have applictions in situations of physical interest ({\it e.g.}, in problems of electron transport in 1D \cite{Giamarchi}).  

A natural extension of our work would be to consider the interacting model, namely the massless Thirring fermion model \cite{Gogolin:2004rp} that is dual to a free compact boson. In this case one can hope that the 
free theory on one side will give insight into the complexity of an interacting theory. The massive Thirring model is dual to a sin-Gordon model, it would be interesting to see if this case could be analyzed using integrability techniques. 

\section*{Acknowledgements}   
It is a pleasure to thank Costas Bachas, Benoit Dou\c cot, Arnaud Fanthomme, Lampros Lamprou, and Rob Myers for some useful conversations at various stages of this work. We would also like to thank Shira Chapman for collaboration at the initial stage of this work and for reading our manuscript and sending us her comments. We would like to thank the Galileo Galilei Institute for Theoretical Physics for hospitality during the workshop “Entanglement in Quantum Systems” where part of this work was carried out. DG acknowledges the LabEx ICFP funding for doctoral studies.
\appendix
\section{Orthogonality of the rotational matrix $M(n)$ for $\alpha\in \mathbb{R}$}\label{app:OrthM}
The orthogonality is independent of the bosonic mode $n$, for simplicity, in this section we will take $n=1$, thus neglect the explicit dependence of $n$ for the matrices. Since $A$ and $B$ are both symmetric, which means
\begin{align}
M M^T = \begin{pmatrix}
A^2 +B^2 & BA-AB\\
AB-BA & A^2 +B^2
\end{pmatrix}.
\end{align}
To prove the orthogonality of $M$ is to show that $(A^2+B^2)_{ik} = \delta_{ik}$ and $(AB-BA)_{ik}=0$, this will be given explicitly in the following part.

\subsection{$(A^2+B^2)_{ik} = \delta_{ik}$}
To start with, $k$ is required to be $k=i+2m (m\in\mathbb{Z})$ since it is easy to see that if $k$ and $i$ differ by an odd number the corresponding entry would give zero. The diagonal entries correspond to the case $m=0$, which are given as 
\begin{align}
(A^2+B^2)_{ii}&= \sum_r A_{i (i+2r)}A_{(i+2r)i} + B_{i (i+2r+1)}B_{(i+2r+1)i}\no\\
&= \sum_r J_{|2r|}^2 + \sum_r J_{|2r+1|}^2 \no\\
&= J_0^2 + 2\sum_{r=1} J_r^2 =1
\end{align}
which is known as Neumann's theorem of Bessel functions. Without loss of generality, we will assume that $m>0$ in the following case for the off-diagonal entries. The calculation shows
\begin{align}
(A^2+B^2)_{ik} &= \sum_{r> m} (-1)^m J_{2r}J_{2(r-m)} +\sum_{r\ge m}  (-1)^m J_{(2r+1)}J_{2(r-m)+1}\no\\
&~~+ \sum_{0\le r\le m}(-1)^m J_{2r}J_{2(m-r)} + \sum_{0\le r<m} (-1)^{m+1}   J_{2r+1}J_{-1-2(r-m)}\no\\
&~~+ \sum_{r<0}(-1)^m (J_{-2r}J_{2(m-r)} + J_{-2r-1}J_{-1-2(r-m)})\no\\
&= (-1)^m \sum_{r\ge 1} J_r J_{2m+r} + (-1)^m\sum_{r=0}^{2m} (-1)^r J_r J_{2m-r} + (-1)^m \sum_{r\ge 1} J_r J_{2m+r}\no\\
&= (-1)^m \left(\sum_{r=0}^{2m} (-1)^r J_r J_{2m-r} +2  \sum_{r\ge 1} J_r J_{2m+r} \right) =0
\end{align}
where in the second step we used the property of Bessel function $J_{-l} = (-1)^l J_l$ and the last step is another theorem of Neumann. Therefore, we have proved that 
\begin{equation}\label{eq:orthoid}
A^2+B^2 =\mathbb{1}
\end{equation}
 is an identity matrix.

\subsection{$(AB-BA)_{ik}=0$}
In this case, the non-zero entries would require $k=2m+1(m \in\mathbb{Z})$. The calculation follows,
\begin{align}
(AB)_{i(i+2m+1)} &= \sum_{r} A_{i(2r+i)}B_{(2r+i)(i+2m+1)}\no\\
&= \sum_r (-1)^r i^{|(2(m-r)+1|-1} J_{2|r|} J_{|2(m-r)+1|}
\end{align}
and 
\begin{align}
(BA)_{i(i+2m+1)} &= \sum_r B_{i(2r+1+i)}A_{(2r+i)(i+2m+1)}\no\\
&= \sum_r (-1)^{m-r} i^{|2r+1|-1} J_{2|m-r|} J_{|2r+1|}\no\\
&= \sum_r (-1)^r i^{|(2(m-r)+1|-1} J_{2|r|} J_{|2(m-r)+1|}
\end{align}
therefore,
\begin{equation}\label{eq:orthozero}
(AB-BA)_{ik}=0.
\end{equation}
Combining \eqref{eq:orthoid} and \eqref{eq:orthozero}, one shows that $M$ is an orthogonal matrix as expected.

\bibliographystyle{JHEP}
\bibliography{bibBC}{}	
\end{document}